\documentclass[aps,prb,twocolumn,showpacs,amsmath,amssymb,footinbib]{revtex4}
\usepackage{graphicx}
\usepackage{amsmath}
\usepackage{enumitem}
\usepackage{natbib}
\usepackage{float}
\usepackage{verbatim}
\usepackage[abs]{overpic}
\usepackage{color}

\begin{document}

\title{Strong effects of weak ac driving in short superconducting junctions}
\author{Roman-Pascal~Riwar, Manuel Houzet, Julia S. Meyer}
\affiliation{Univ. Grenoble Alpes, INAC-SPSMS, F-38000 Grenoble, France;\\
CEA, INAC-SPSMS, F-38000 Grenoble, France.
}
\author{Yuli~V.~Nazarov}
\affiliation{Kavli Institute of NanoScience, Delft University of Technology, Lorentzweg 1, NL-2628 CJ, Delft, The Netherlands.}


\begin{abstract}
We study a short superconducting junction subject to a dc and ac phase bias. The ac modulation changes the occupation of the Andreev bound states formed at the constriction by transitions between bound states and the continuum. In a short junction, the non-equilibrium Andreev bound state population may relax through processes that conserve parity of the occupation number on the same bound state and processes that do not conserve it. We argue that the parity conserving processes occur on a much faster time scale. In this case, even a weak driving may lead to a large deviation of the supercurrent from its equilibrium value. We show that this effect is accompanied by a quasiparticle current which may lead to a measurable charge imbalance in the vicinity of the junction.  Furthermore, we study the time evolution of the supercurrent after switching off the ac drive. On a time scale where parity relaxation is negligible, the supercurrent relaxes to a stationary non-equilibrium state. Finally, we briefly outline the regime of ultraweak driving where the ac-induced processes occur on a time scale comparable to parity relaxation.

\end{abstract}

\maketitle

\section{Introduction}

The ac properties of superconducting junctions have been under extensive scrutiny, both theoretically~\cite{Josephson1962,Virtanen2010,Kos2013} and experimentally~\cite{LevensonFalk2011,Chen2011,Fuechsle2009}, due to a manifold of reasons, such as the possibility of a microwave-induced enhancement of the critical current or generally for the study of high-frequency superconducting circuits.

In general, an ac irradiation gives rise to a non-equilibrium excitation of the superconductor, and thus leads to a finite quasiparticle population.
The presence of quasiparticles remains a challenge inherent to the superconductor~\cite{Catelani2011}, because the trapping of quasiparticles inside the device, commonly referred to as quasiparticle poisoning, may lead to decoherence, as has been discussed in various setups~\cite{Martinis2009,Lenander2011,Fu2009,vanHeck2011,Rainis2012,Rajauria2012,Riste2013,Wenner2013,LevensonFalk2014}. Electronic cooling with the aim to reduce the number of quasiparticle excitations has been shown to lack efficiency~\cite{Leivo1996,Rajauria2007}, which makes the control of the quasiparticle population an important present-day topic.

In this paper, we consider multichannel superconducting junctions, subject to an ac phase drive. Importantly, we focus on the limit of short junctions, which means that the different channels may be regarded as independent.
Then, a finite phase bias $\phi$ gives rise to an Andreev bound state (ABS) for each channel~\cite{Andreev1964,QuantumTransport}. As the ABS supports a supercurrent, it offers the possibility of the direct experimental observation of the bound state occupation~\cite{BretheauRPL}. A non-equilibrium driving of the phase bias~\cite{Bergeret2011,Kos2013,Olivares2014,Zazunov2014,Bretheau2014} was experimentally deployed for the spectroscopy of Andreev levels~\cite{Bretheau2013,Bretheau2013b}.

A weak ac drive couples the bound state with the continuum via two processes. A quasiparticle may be directly promoted from a bound state to the continuous spectrum, which gives rise to the ionization rate $\Gamma_\text{I}$. Alternatively, a Cooper pair can be broken, whereby a quasiparticle fills the bound state and a second one goes to the continuum, described by the refill rate $\Gamma_\text{R}$. 
Both of these processes change the parity of the bound state occupation number.

By contrast, at sufficiently low temperatures the dominant relaxation mechanism in a short junction does not involve the continuum, and thus preserves parity~\cite{BretheauRPL}. Namely, there is a parity-conserving relaxation process of the annihilation of two bound quasiparticles on the same level due to environment induced phase fluctuations, giving rise to the annihilation rate $\Gamma_\text{A}$.

In Ref.~\onlinecite{Riwar2014} we showed in the single channel case that
due to the suppressed parity relaxation, even a weak modulation establishes a significant non-equilibrium bound state population. Furthermore, we demonstrated that the ac driving also leads to a non-equilibrium charge accumulation of the excess quasiparticles near the constriction, due to a charge asymmetry of the ac-induced refill and ionization processes. 

In the present paper we extend our considerations to a short multichannel superconducting junction~\cite{Virtanen2010,Kos2013} with a conductance $G$ much larger than the conductance quantum $G_{Q}=e^2/\pi\hbar$. In this case the ABS form a quasicontinuous spectrum. The short junction limit implies that we can disregard any energy dependence of the transmission through the junction which occurs at the energy scale of the Thouless energy, $E_\text{Th}$, which we assume to be much larger than the superconducting gap $\Delta$. In this case, the ABS corresponding to different channels are orthogonal, and the dominant relaxation mechanism is still the annihilation of two quasiparticles in the \textit{same} bound state. Consequently, the parity of each individual ABS relaxes very slowly. Possible parity relaxation processes include the annihilation of two quasiparticles in different ABS, and the relaxation of a quasiparticle by a transition to a lower lying ABS. However, the corresponding rates of these processes would be smaller by the factor $\Delta^2/E_\text{Th}^2$. Therefore, for a sufficiently strong driving, these processes can be disregarded, leaving us with the single channel annihilation process only, occurring with the rate $\Gamma_\text{A}$. As a consequence, the contributions of each channel simply add up. Thus, already a weak ac modulation, $\Gamma_\text{I,R}\ll\Delta$ (yet $\Gamma_\text{I,R}$ much larger than the parity relaxation), can lead to a strong ac-induced change of the population even for the multiterminal junction.

The non-equilibrium population of ABS manifests itself as a strong modification of the superconducting current through the junction with respect to its equilibrium value. As an interesting peculiarity, we discuss the reversal of the sign of the supercurrent near the stationary phase $\phi=\pi$. A similar feature has been reported in long superconducting junctions~\cite{Fuechsle2009,Voutilainen2012}.

In analogy to the single-channel case~\cite{Riwar2014}, the charge imbalance of the ionization and refill processes leads to an extra dissipative quasiparticle current, $e\dot{q}$, where $\dot{q}$ signifies the ac-induced charge transfer. 
Such a charge imbalance can be measured by the standard setup of a normal metal-superconducting (N-S) contact~\cite{Langenberg1986,Tinkham1972b,Tinkham1972}, by use of which the decay of the non-equilibrium charge was investigated experimentally~\cite{Hubler2010,Golikova2014}. 
Either the quasiparticle current $e\dot{q}$ is collected by a sufficiently conducting normal metal contact in the vicinity of the superconducting junction, or the N-S contact is used to measure the voltage signal due to the non-equilibrium charge accumulation. Importantly, in the multichannel case the quasiparticle current can reach a large value, $\dot{q}\sim(\delta\phi)^2I_\text{c}$, where $I_\text{c}$ is the critical supercurrent of the junction. The multichannel junction thus facilitates the experimental observation of this effect.

The slow parity relaxation motivates us to consider the time evolution of the supercurrent after switching off the ac driving, within the time scale set by $\Gamma_\text{A}^{-1}$. As before, the superconducting current provides a direct signal of the time evolution of the quasiparticle occupation. On the time scale $\Gamma_\text{A}^{-1}$, the supercurrent does not relax to its equilibrium value. In fact, the supercurrent is lowered in magnitude due to the persisting singly occupied bound states. After an initial exponential onset of the decay, we find that in a diffusive junction, the time evolution of the current asymptotically approaches $\sim1/\sqrt{t}$. This is due to the long lifetime of the doubly occupied bound states of the highly transparent channels, for which $\Gamma_\text{A}\rightarrow 0$. To describe thermalization at even longer times, parity relaxation mechanisms have to be taken into account.

Namely, in the absence of a parity relaxation mechanism, a single bound quasiparticle would be trapped forever, unless being affected by the ac modulation. In reality, parity relaxation can occur for instance by annihilation with a delocalized thermally excited quasiparticle, or due to the above mentioned interchannel processes. Owing to the complexity of the possible processes, we introduce a phenomenological relaxation rate $\Gamma_\text{P}\ll\Gamma_\text{A}$ without providing any microscopic model for it. This mechanism becomes important when $\Gamma_\text{I,R}\sim\Gamma_\text{P}$, i.e., in the ultraweak driving regime. When $\Gamma_\text{I,R}\ll\Gamma_\text{P}$, the supercurrent approaches its equilibrium value. However, charge imbalance effects may survive, depending on the origin of parity relaxation. 
Note that, a common theoretical description of the relaxation in superconducting junctions (see, e.g., Ref.~\onlinecite{Bergeret2011}) involves the artificial assumption of a small but finite quasiparticle density of states within the superconducting gap. This model does not provide a different decay process for even and odd occupation. We believe that our approach better reflects physical reality~\cite{BretheauRPL}.

The structure of this paper is as follows. In Sec.~\ref{sec_dynamics} we outline the model and describe the dynamics of the bound state population in presence of the ac drive, when $\Gamma_\text{I,R}\gg\Gamma_\text{P}$. In Sec.~\ref{sec_supercurrent} we discuss the change of the supercurrent when modifying the ABS population as a result of the ac drive. In Sec.~\ref{sec_charge_imbalance} we present the results for the charge imbalance created in the contacts due to the ac-induced ionization and refill processes of the ABS. In Sec.~\ref{sec_supercurrent_time} we consider the time evolution of the supercurrent after switching off the ac modulation, on the time scale of the lifetime of the doubly occupied bound states. Finally, we discuss the regime of ultraweak ac drive, when the rates $\Gamma_\text{I,R}$ are comparable with $\Gamma_\text{P}$ in Sec.~\ref{sec_parity}. We conclude our findings in Sec.~\ref{sec_conclusions}.

\section{Dynamics of ac driven multichannel junction}\label{sec_dynamics}

\begin{figure}
\includegraphics[width=1\columnwidth]{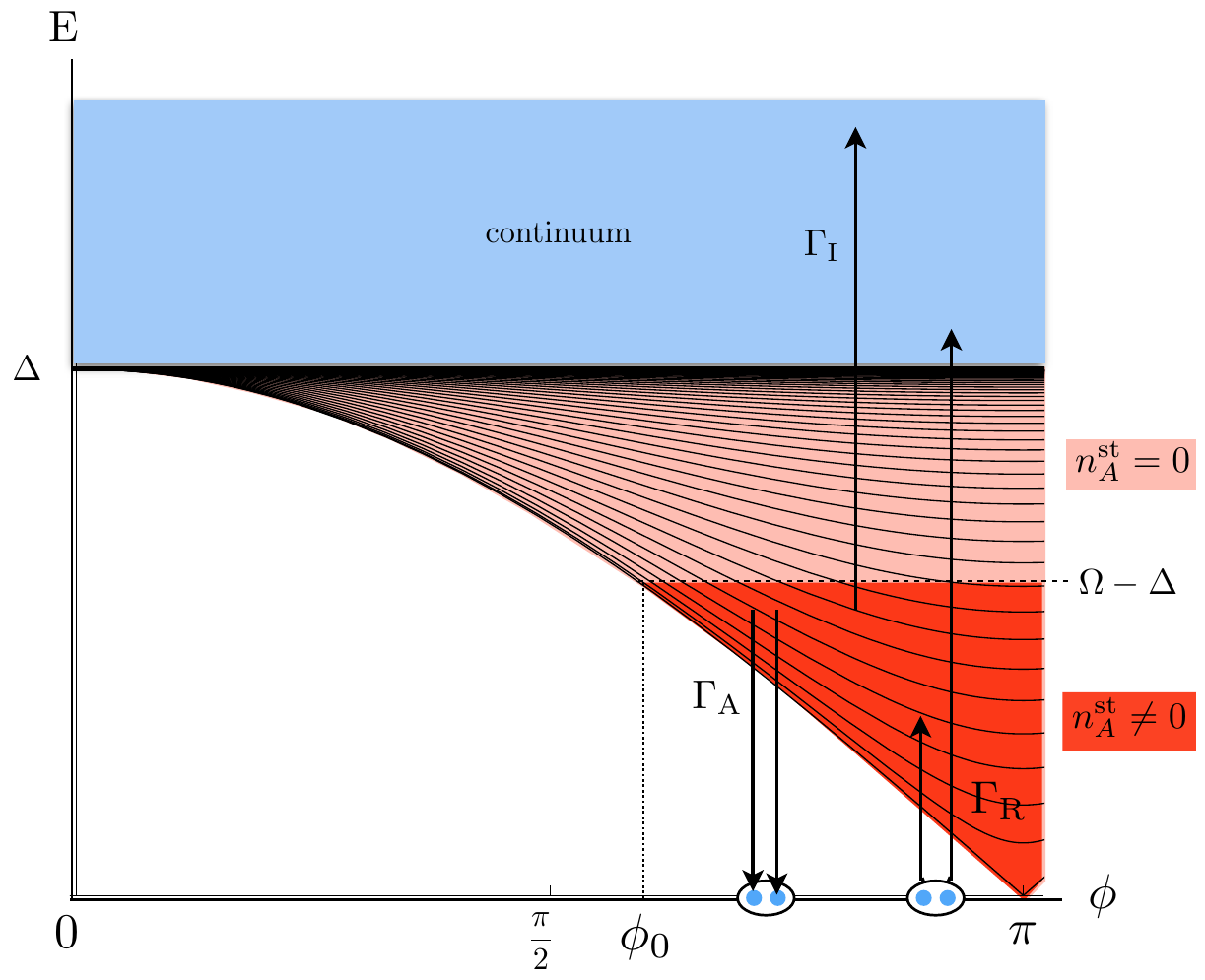}
\caption{Energy diagram of the multichannel weak link as a function of the stationary phase difference $\phi$. The collection of red curves indicates the Andreev bound state energies below the gap. The blue area represents the continuum of free quasiparticle states above the gap. The arrows show the processes that change the ABS occupations. The ionization process, occurring with rate $\Gamma_\text{I}$, promotes a quasiparticle in a bound state to the continuum. The refill process ($\Gamma_\text{R}$) breaks a Cooper pair, and occupies a bound state while releasing a second quasiparticle into the continuum. In addition to these ac-induced processes, phase fluctuations induce the annihilation of two quasiparticles in the same bound state, the rate of which is $\Gamma_\text{A}$. Channels with a bound state energy above the threshold $\Omega-\Delta$ are empty, while states below this threshold can be populated, $n_A^\text{st}\neq0$, due to the ac-induced refill process, $\Gamma_\text{R}$.}\label{fig_energy_diagram}
\end{figure}

We aim to study the manipulation of the bound state occupation in a weak superconducting link with many channels, due to an ac drive of the phase bias. In the short junction limit, within each channel a bound state with energy $E_A(T)=\Delta\sqrt{1-T\sin^2\frac{\phi}{2}}$ is formed, where $\phi$ is the phase bias and $T$ is the transparency of an individual channel. Thus, each of the channels gives rise to a contribution to the supercurrent, the sum of which is
\begin{equation}\label{eq_Is}
I_{S}=-2e\int_{0}^{1}dT\rho\left(T\right)\frac{\partial E_{A}}{\partial\phi}\left(T\right)\left[1-n_A(T)\right]\ .
\end{equation}
Here, $n_A(T)$ is the occupation of the ABS in the channel with transmission $T$. 
At a low temperature, all ABS are empty and the equilibrium current is simply given as $I_S^\text{eq}\equiv I_S(n_A=0)$. The values of the channels' transparencies occur according to the distribution function $\rho\left(T\right)$, which depends on the details of the junction. Here, we concentrate on two examples, a diffusive junction and a double tunnel junction. The Dorokhov distribution for a diffusive junction is~\cite{QuantumTransport},
\begin{equation}\label{eq_rho_D}
\rho_\text{D}(T)=\frac{1}{2}\frac{G}{G_{Q}}\frac{1}{T\sqrt{1-T}}\ ,
\end{equation}
where $G$ is the Drude conductance. In the diffusive junction, there is both a high concentration of low transparency channels, $T\ll 1$, and a high concentration of almost transparent channels, $T\sim 1$. The sketch in Fig.~\ref{fig_energy_diagram} schematically represents the corresponding distribution of bound state energies for the multichannel wire (red curves). 

The distribution for a double junction is given as~\cite{QuantumTransport}
\begin{equation}\label{eq_rho_dj}
\rho_\text{dj}\left(T\right)=\frac{1}{\pi}\frac{G}{G_{Q}}\frac{1}{T^{3/2}\sqrt{T_{c}-T}}\theta\left(T_c-T\right)\ ,
\end{equation}
where the Heaviside theta function $\theta$ indicates that $\rho_\text{dj}$ in non-zero only for $T\leq T_c$. The total conductance is $G=G_1G_2/(G_1+G_2)$, where $G_{1,2}$ stand for the conductances of the individual junctions. Here, there is likewise a high concentration of small transparencies. However, in the double junction in general the high transparency channels occur up to the critical $T_c=4G_1G_2/(G_1+G_2)^2$. Hence $T_c<1$, unless the junction is symmetric, $G_1=G_2$.

\begin{figure}
\includegraphics[width=0.65\columnwidth]{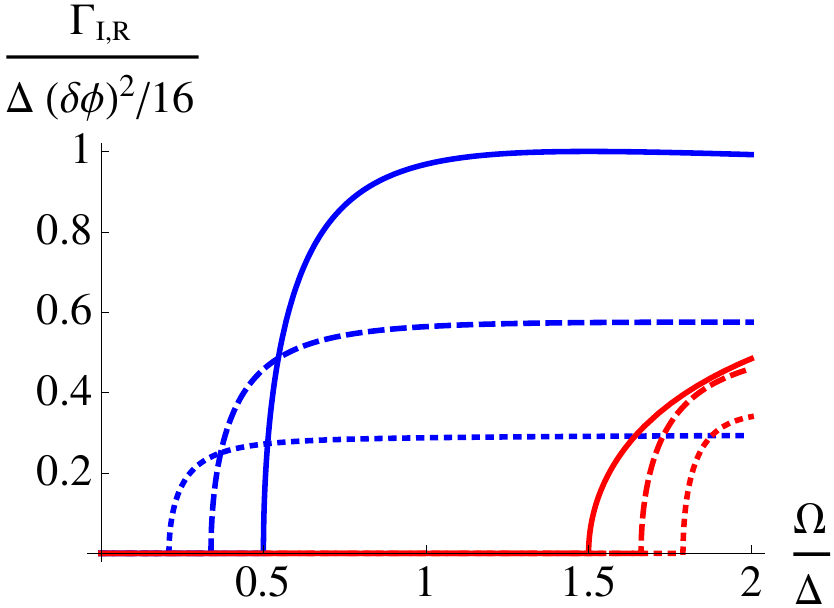}
\caption{The ionization (blue) and refill (red) rates as functions of the driving frequency $\Omega$ for the transparencies $T=0.5,0.75,1$ (dotted, dashed, solid). The phase bias is $\phi=2\pi/3$.}
\label{fig_Gamma_I_R}
\end{figure}

We consider the change of the occupation of the ABS of an ac phase driving, $\phi(t)=\phi+\delta\phi\cos(\Omega t)$, with the driving frequency $\Omega$ and the \textit{weak} driving amplitude $\delta\phi\ll 1$. In addition, we restrict our considerations to a driving frequency $\Omega\leq 2\Delta$. A driving frequency above $2\Delta$ may result in a large non-equilibrium quasiparticle population in the leads. In the short junction limit, the driving does not couple the channels. Therefore we may treat each channel separately. The ABS occupation of a channel with transparency $T$, $n_A(T)$, may be changed due to an \textit{ionization} process whereby a bound state quasiparticle is promoted to the continuum or the ABS may be \textit{refilled} when a Cooper pair is broken to create a quasiparticle in both a bound state and the continuum, see Fig.~\ref{fig_energy_diagram}. The corresponding rates are calculated via Fermi's golden rule (see Ref.~\onlinecite{Riwar2014}),
\begin{align}\label{eq_Gamma_I_R}
\Gamma_\text{I,R}(T)=&\frac{T(\delta\phi)^2}{16}\theta(\Omega\pm E_A-\Delta)\frac{\sqrt{\Delta^2-E_A^2}}{E_A}\\\nonumber &\times\sqrt{(\Omega\pm E_A)^2-\Delta^2}\frac{E_A\Omega\pm\Delta(\cos\phi+1)}{(\Omega\pm E_A)^2-E_A^2}\ .
\end{align}
Note that in total, the ionization and refill rates scale as $\Delta(\delta\phi)^2$, which is much smaller than $\Delta$, the characteristic energy scale of the bound states. We recognize that the minimal threshold for an ionization/refill process is $\Omega>\Delta\mp E_A$ (see also Fig.~\ref{fig_Gamma_I_R}). As for the stationary ABS population, it is the refill process that defines the threshold for a finite population of bound states, see also Fig.~\ref{fig_energy_diagram}, which is why we consider irradiation frequencies above $\Omega>\Delta$. As the bound state energy is a function of $\phi$, there is a minimal stationary phase $\phi_0$ at which bound states begin to be populated, namely
\begin{equation}\label{eq_phi_0}
\phi_0=2\arcsin\left(\sqrt{\frac{\Omega(2\Delta-\Omega)}{\Delta^2T_\text{max}}}\right)\ ,
\end{equation}
where $T_\text{max}=1$ for the diffusive junction, and $T_\text{max}=T_c$ for the double tunnel junction. As indicated in Fig.~\ref{fig_energy_diagram} the window of occupied bound states is the one with lower energy, and thus involves the channels with higher transparency. In the limit $\Omega\rightarrow2\Delta$, the threshold phase goes to zero, $\phi_0\rightarrow0$. 

Note that for $\Omega$ in the vicinity of $2E_A$, the ac modulation can also drive coherent Rabi oscillations between the zero and doubly occupied state. The interval of energies surrounding $2E_A$, where such processes can in general not be neglected, scales as $\delta\phi\Delta$. The interval of all bound states that contribute to the supercurrent scales, however, with $\Delta$ and is therefore much larger. This is why one may safely neglect the small fraction of channels that might be affected by ac driven Rabi oscillations.

Our model includes fluctuations of the phase due to the coupling to a noisy environment, which gives rise to an annihilation process whereby a doubly occupied ABS is completely emptied. The rates of such annihilation events are
\begin{equation}
\Gamma_\text{A}(T)=\frac{S_\phi(2E_A)}{4}\left(1-\frac{E_A^2}{\Delta^2}\right)\left(\Delta^2-E_A^2-4\left[\frac{\partial E_A}{\partial\phi}\right]^2\right)\ ,
\end{equation}
where $S_\phi(\omega)$ stands for the phase noise spectrum. For the environment temperature much smaller than the frequency $\omega$, it may be related to the impedance of the external circuit~\cite{Ingold1992}, $Z(\omega)$, as $S_\phi(\omega)=4\pi G_{Q}Z(\omega)/\omega$. For the sake of simplicity we assume an ohmic environment, such that $S_\phi(\omega)\equiv (\phi_q)^2/\omega$, where $(\phi_q)^2$ expresses the phase noise power.

We may now formulate the rate equations for each individual channel with transparency $T$, that define the time evolution of the bound state occupation of that
channel,
\begin{align}
\dot{P}_{0}\left(T\right) =& -2\Gamma_{\text{R}}\left(T\right)P_{0}\left(T\right)+\Gamma_{\text{I}}\left(T\right)
P_{1}\left(T\right)\\\nonumber
&+\Gamma_{\text{A}}\left(T\right)P_{2}\left(T\right)\ ,\\
\dot{P}_{1}\left(T\right) =&  -\left[\Gamma_{\text{R}}\left(T\right)+\Gamma_{\text{I}}\left(T\right)\right]P_{1}\left(T\right)\\\nonumber
&+2\Gamma_{\text{R}}\left(T\right)P_{0}\left(T\right)+2\Gamma_{\text{I}}\left(T\right)P_{2}\left(T\right)\ ,
\end{align}
where the probabilities to have zero, one, and double occupation, $P_0$, $P_1$, and $P_2$, satisfy $P_0+P_1+P_2=1$. The total occupation is given as $n_A=P_1+2P_2$.

The stationary solutions are readily computed as
\begin{align}\label{eq_stationary_P0}
P_{0}^{\text{st}}\left(T\right) =&\frac{2\Gamma_{\text{I}}^{2}\left(T\right)+\Gamma_{\text{A}}\left(T\right)
\left[\Gamma_{\text{I}}\left(T\right)+\Gamma_{\text{R}}\left(T\right)\right]}{2\left[\Gamma_\text{R}(T)+\Gamma_\text{I}(T)\right]^2+\Gamma_\text{A}(T)\left[3\Gamma_\text{R}(T)+\Gamma_\text{I}(T)\right]}\ ,\\ \label{eq_stationary_P1}
P_{1}^{\text{st}}\left(T\right) =&\frac{4\Gamma_{\text{R}}\left(T\right)\Gamma_{\text{I}}\left(T\right)+2\Gamma_{\text{A}}\left(T\right)\Gamma_{\text{R}}\left(T\right)}{2\left[\Gamma_\text{R}(T)+\Gamma_\text{I}(T)\right]^2+\Gamma_\text{A}(T)\left[3\Gamma_\text{R}(T)+\Gamma_\text{I}(T)\right]}\ .
\end{align}
There are two limiting cases where the stationary occupations no longer depend on $\Gamma_\text{A}$. When $\Gamma_\text{I,R}\gg\Gamma_\text{A}$, corresponding to $(\delta\phi)^2\gg(\phi_q)^2$, the stationary probabilities reduce to 
\begin{align}\label{eq_stationary_P0_strong}
P_{0}^{\text{st}}\left(T\right)&=\frac{\Gamma_{\text{I}}^{2}\left(T\right)}{\left[\Gamma_{\text{I}}\left(T\right)+\Gamma_{\text{R}}\left(T\right)\right]^{2}}\ ,\\ \label{eq_stationary_P1_strong}
P_{1}^{\text{st}}\left(T\right)&=\frac{2\Gamma_{\text{I}}\left(T\right)\Gamma_{\text{R}}\left(T\right)}{\left[\Gamma_{\text{I}}\left(T\right)+\Gamma_{\text{R}}\left(T\right)\right]^{2}}\ .
\end{align}
In the opposite limit, when $\Gamma_\text{I,R}\ll\Gamma_\text{A}$, i.e., $(\delta\phi)^2\ll(\phi_q)^2$, 
\begin{align}\label{eq_stationary_P0_weak}
P_{0}^{\text{st}}\left(T\right) =&\frac{\Gamma_{\text{I}}\left(T\right)+\Gamma_{\text{R}}\left(T\right)}{3\Gamma_\text{R}(T)+\Gamma_\text{I}(T)}\ ,\\ \label{eq_stationary_P1_weak}
P_{1}^{\text{st}}\left(T\right) =&\frac{2\Gamma_{\text{R}}\left(T\right)}{3\Gamma_\text{R}(T)+\Gamma_\text{I}(T)}\ .
\end{align}
whereas the double occupation is suppressed, $P_2^\text{st}=0$.

\section{Supercurrent signal of the irradiated junction}\label{sec_supercurrent}

As shown in the previous section, by applying an ac drive it is possible to control the bound state occupation, which in turn changes the supercurrent. In this section we consider the case when the ac drive is permanent, such that the stationary occupation is reached.  Then, the supercurrent is obtained by inserting $n_A^\text{st}=P_1^\text{st}+2P_2^\text{st}$ from Eqs.~\eqref{eq_stationary_P0} and~\eqref{eq_stationary_P1} into Eq.~\eqref{eq_Is}. In the here considered regime of negligible parity relaxation, $\Gamma_\text{I,R}\gg\Gamma_\text{P}$, the stationary occupations do not scale with the driving power. Therefore, the supercurrent may strongly deviate from its equilibrium value, in contrast with the considerations of Ref.~\onlinecite{Bergeret2011}.

\begin{figure*}
(a)\includegraphics[width=0.65\columnwidth]{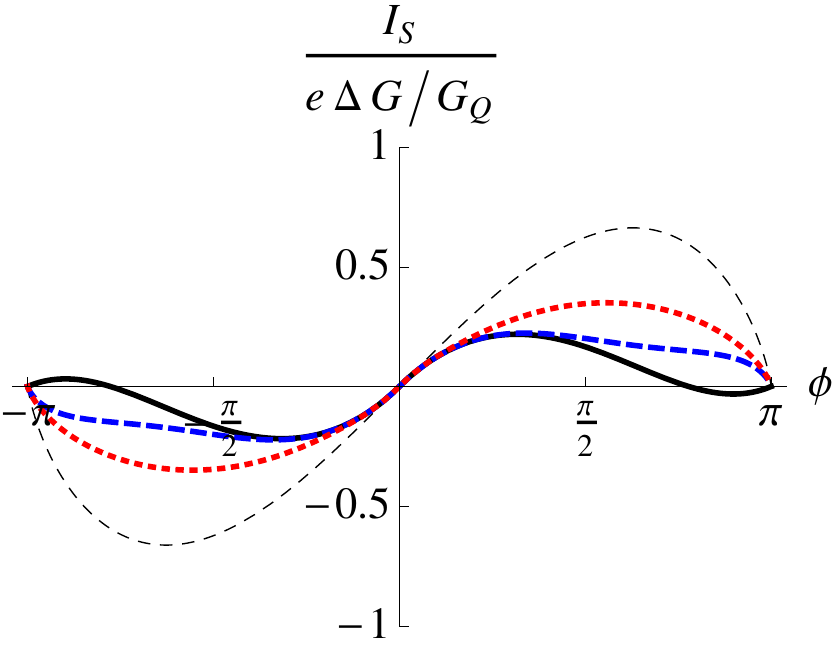}(b)\includegraphics[width=0.65\columnwidth]{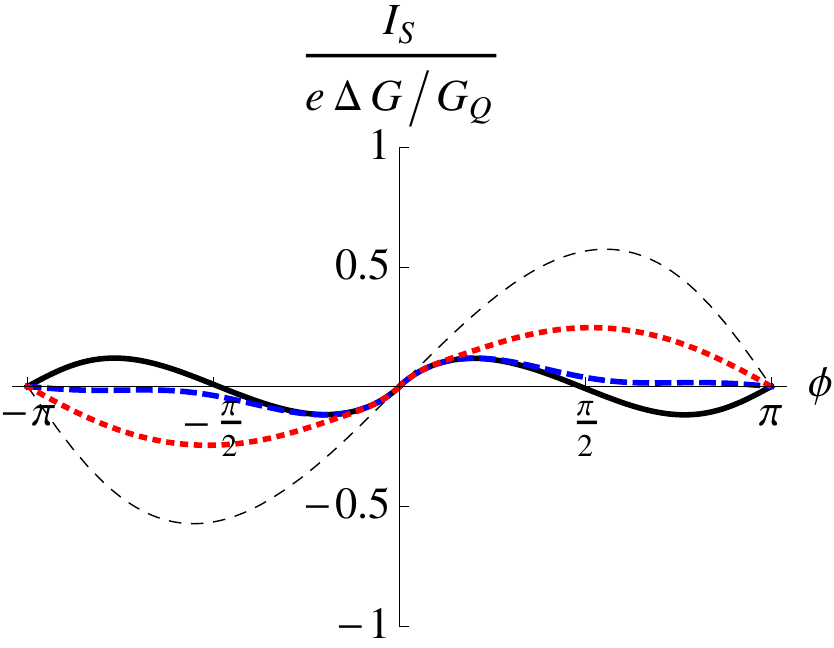}

(c)\includegraphics[width=0.65\columnwidth]{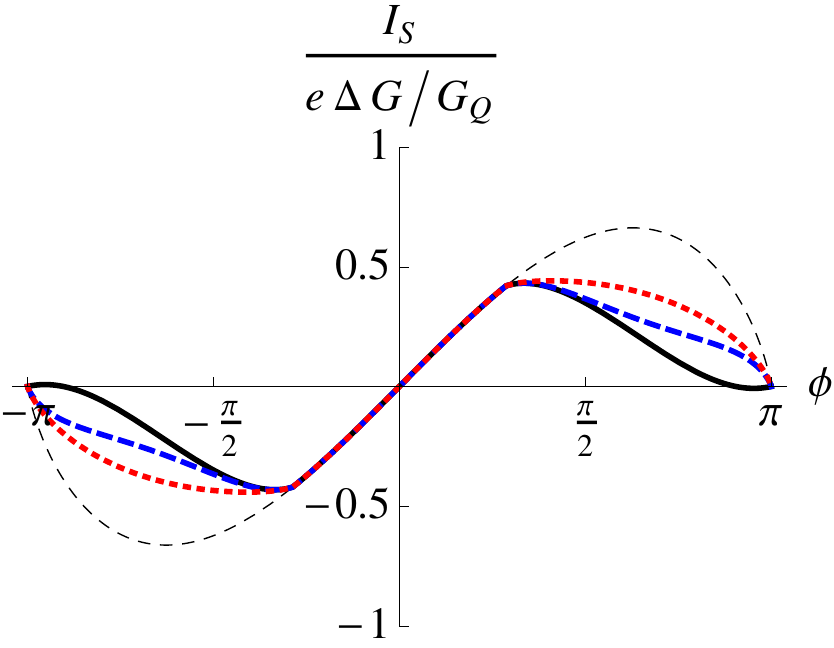}(d)\includegraphics[width=0.65\columnwidth]{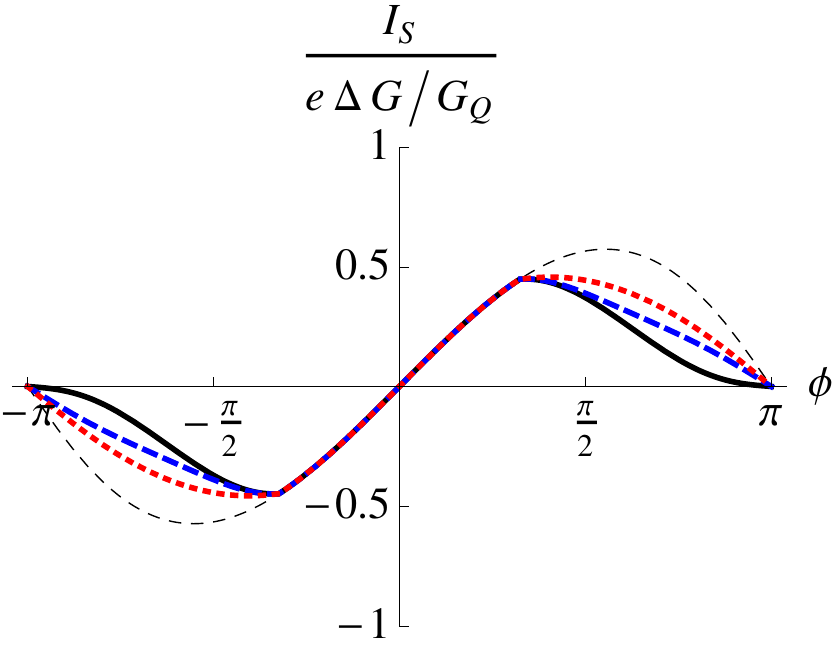}

(e)\includegraphics[width=0.65\columnwidth]{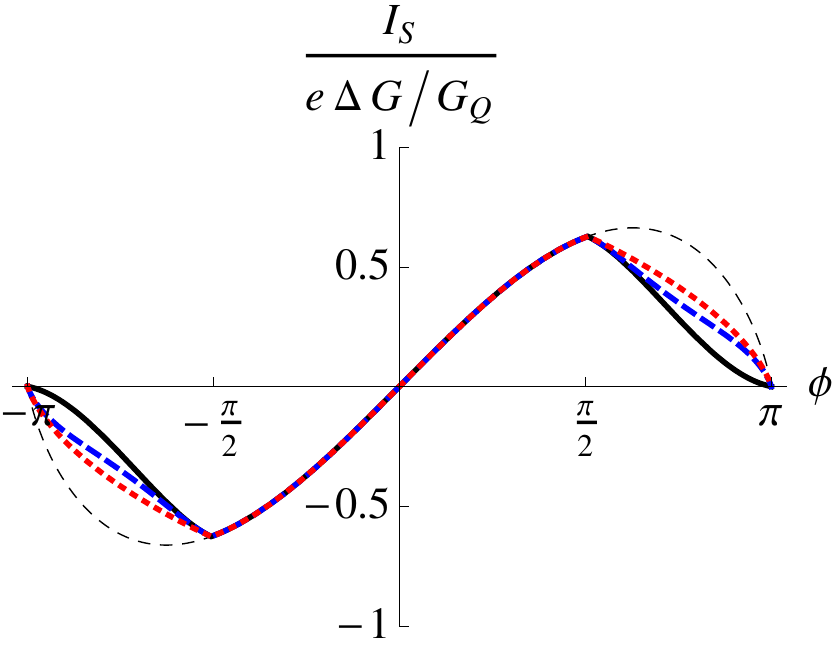}(f)\includegraphics[width=0.65\columnwidth]{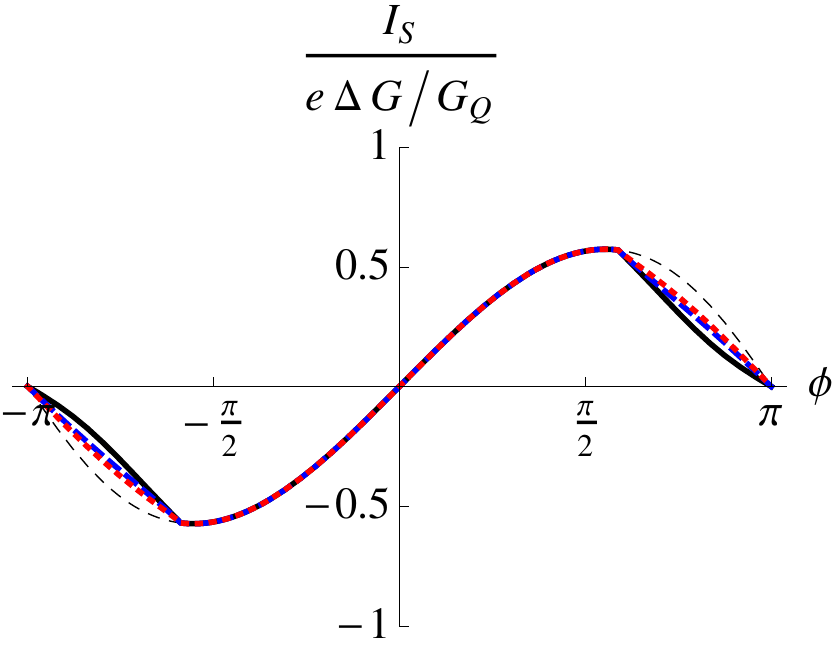}

\caption{The supercurrent $I_{S}$ as a function of the stationary phase $\phi$ for a diffusive junction (a,c,e) and a double junction with $T_c=0.8$ (b,d,f). We show the supercurrent for $\delta\phi/\phi_{q}=0.1$ (dotted line),
$\delta\phi/\phi_{q}=1$ (thick dashed line), and $\delta\phi/\phi_{q}=10$
(solid line). The equilibrium supercurrent in the absence of an ac drive is shown as the thin dashed line. The driving frequencies are $\Omega=2\Delta$ (a and b), $\Omega=1.9\Delta$ (c and d), and $\Omega=1.7\Delta$ (e and f).}
\label{fig_Is}
\end{figure*}

Figure~\ref{fig_Is} shows the current-phase relation of the driven junction. In general, the driving leads to a suppression of the supercurrent as the bound state occupation is increased, $n_\text{A}>0$. We see that the ac-induced change of the current is comparable to the equilibrium supercurrent.
For $\Omega=2\Delta$ (Figs.~\ref{fig_Is}a and~b), the supercurrent is affected by the ac driving for all stationary phases, whereas for $\Omega<2\Delta$ (Figs.~\ref{fig_Is}c-f) we clearly see that the supercurrent changes only when the absolute value of the phase exceeds the threshold phase $\phi_0$ [see Eq.~\eqref{eq_phi_0}].

The onset of the ac driven change of the supercurrent at a finite $\phi_0$ gives rise to a cusp. This cusp originates from the large density of channels with a transparency close to $T_\text{max}$, which get occupied first near the threshold phase. This feature is analyzed in more detail in Appendix~\ref{app_cusp}.

\begin{figure*}
(a)\includegraphics[width=0.65\columnwidth]{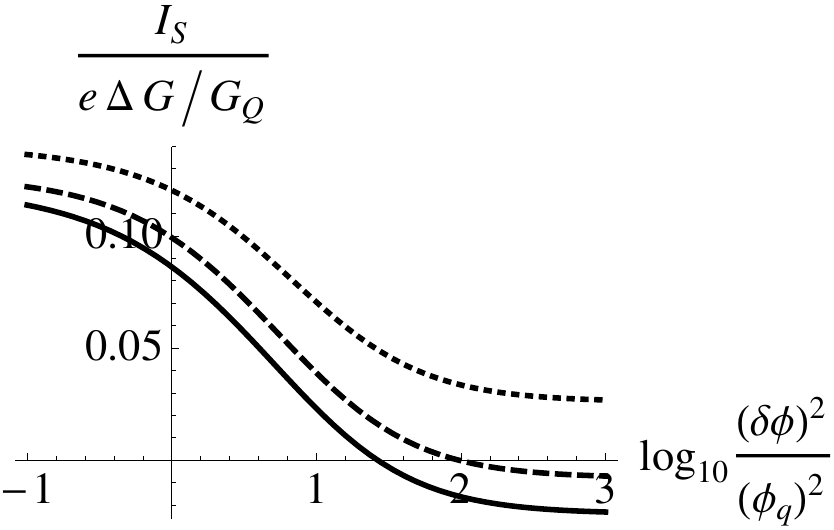}\hspace{7mm}(b)\includegraphics[width=0.65\columnwidth]{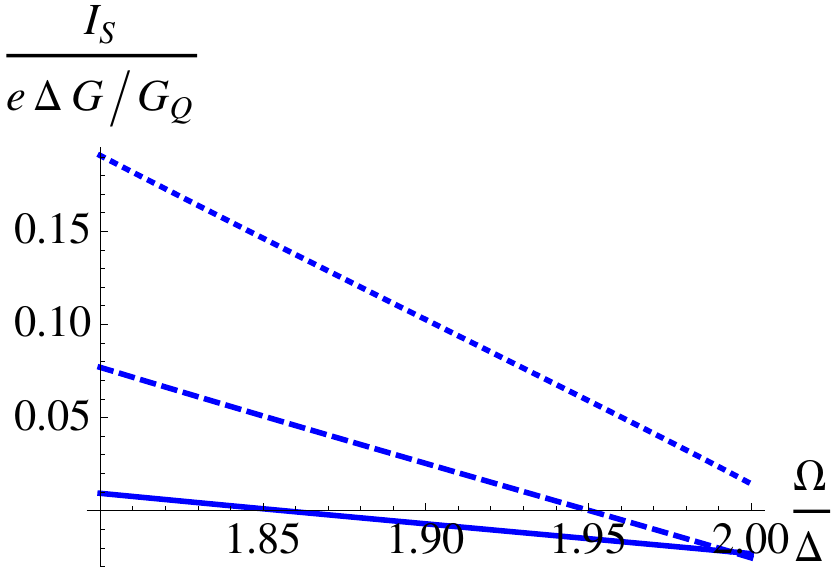}

(c)\includegraphics[width=0.65\columnwidth]{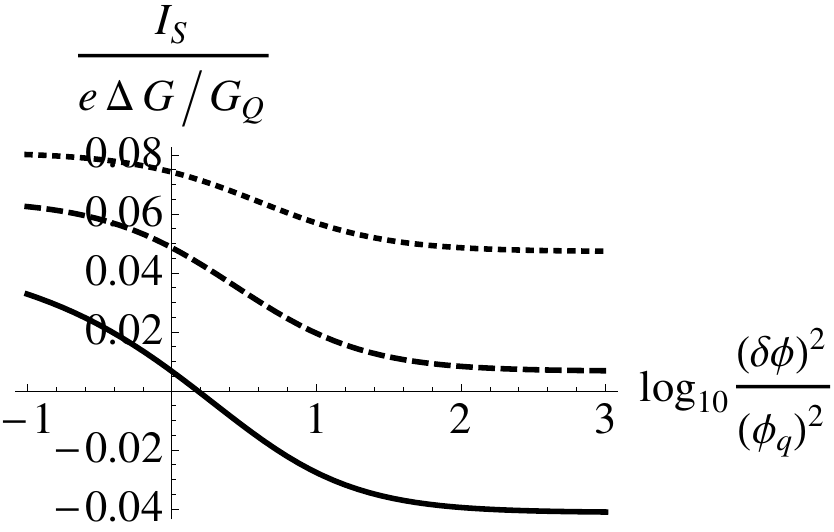}\hspace{7mm}(d)\includegraphics[width=0.65\columnwidth]{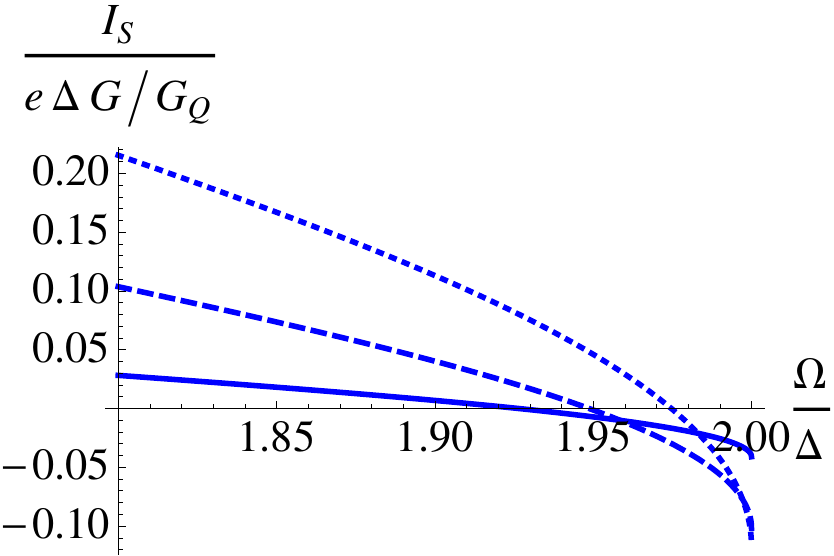}

\caption{The supercurrent $I_{S}$ as a function of driving power (a and c) and as a function of driving frequency (b and d). Figures a and b show the results for the diffusive junction, while in c and d we display the result of the double junction. In a and c the curves are shown for $\phi=0.95\pi$ and the driving frequencies $\Omega={1.7\Delta,1.9\Delta,2\Delta}$ (dotted, dashed, solid). In b and d the curves are drawn for the driving power $(\delta\phi)^2=1000(\phi_q)^2$ and for the different phases $\phi={0.75\pi,0.85\pi,0.95\pi}$ (dotted, dashed, solid).}
\label{fig_Is_inversion}
\end{figure*}

The supercurrent signal changes its behaviour with respect to the driving strength. In Figs.~\ref{fig_Is}a and~b (and very faintly in c), one may observe that the supercurrent can change sign with respect to the equilibrium current, for a stationary phase bias close to $\pi$, when $\Gamma_\text{I,R}\gg\Gamma_\text{A}$. 
We can explain the occurence of the sign reversal as follows. Consulting the general expression for the supercurrent in Eq.~\eqref{eq_Is}, we realize that such an inversion of the supercurrent can only occur if two conditions are fulfilled. Firstly, the driving must be strong enough to occupy the bound states with more than one quasiparticle, $n_A>1$. This requires $\Gamma_\text{I,R}\gg\Gamma_\text{A}$, or equivalently $(\delta\phi)^2\ll(\phi_q)^2$. Secondly, a sufficient number of channels must be affected by the driving, such that the effect survives in the total supercurrent, summed over all channels. This means that the driving frequency should be sufficiently large so that the contribution of channels with $E_A<\Omega-\Delta$, which acquire a population $n_A>1$, overcomes that of the remaining channels. In Fig.~\ref{fig_Is_inversion}a and c, we show the supercurrent $I_S$ as a function of driving power, for a stationary phase $\phi$ close to $\pi$. We see indeed that the supercurrent inversion can occur only for $(\delta\phi)^2\gg(\phi_q)^2$, provided that $\Omega$ is sufficiently high.

A similar sign inversion of the supercurrent has been reported experimentally in Ref.~\onlinecite{Fuechsle2009} and discussed theoretically in Ref.~\onlinecite{Voutilainen2012}. We have to note that in said works, the authors studied a \textit{long} superconducting junction, whereas we predict this effect for the model of a \textit{short} junction. This is why a direct comparison is not possible.

Note that for driving frequencies close to $\Omega\approx2\Delta$, there is a difference in the behaviour of the supercurrent for the diffusive and double junction. Namely, it can be shown (see Appendix~\ref{app_I_S_high_frequency}) that the supercurrent changes linearly with $\delta\Omega= 2\Delta-\Omega$ for the diffusive case,
\begin{equation}\label{eq_Is_Omega_rho_D}
I_S\approx \left.I_S\right|_{\Omega=2\Delta}-2e\frac{G}{G_{Q}}\frac{\partial^2E_A}{\partial\phi\partial T}(0)\left.n_A^\text{st}(0)\right|_{\Omega=2\Delta}\frac{\delta\Omega}{\sin^2\frac{\phi}{2}\Delta}\ .
\end{equation}
Likewise we can show that the supercurrent in the double junction is approximated as,
\begin{equation}\label{eq_Is_Omega_rho_dj}
I_S\approx\left.I_S\right|_{\Omega=2\Delta}-e\frac{4\sqrt{2}G}{\pi G_{Q}}\frac{\partial^2E_A}{\partial\phi\partial T}(0)\left.n_A^\text{st}(0)\right|_{\Omega=2\Delta}\sqrt{\frac{\delta\Omega}{\sin^2\frac{\phi}{2}\Delta}}\ ,
\end{equation}
and therefore scales with the square root of $\delta\Omega$. This different scaling behaviour with respect to the driving frequency is depicted in Figs.~\ref{fig_Is_inversion}b and~d. Let us provide a qualitative picture. At a driving frequency close to $2\Delta$, it is the filling of the low $T$ channels that determines the behaviour of the supercurrent. The characteristic behaviour of the supercurrent for the two different junctions is simply due to the respective behaviour of the density of the low $T$ channels. Namely, for the diffusive junction $\rho_\text{D}\sim T^{-1}$, while for the double junction $\rho_\text{dj}\sim T^{-3/2}$. The concentration of low $T$ channels in the double junction increases faster than in the diffusive junction. This gives rise to the more pronounced change of the supercurrent in the double junction with respect to the driving frequency. Consequently, the supercurrent is directly providing information about the channel distribution of the junction for low $T$. Note that for $\phi$ close to zero, the high energy channels are equally important in the filling process for frequencies close to $2\Delta$, and therefore, the above argument does not work.

\section{Charge imbalance}\label{sec_charge_imbalance}
In the previous section, we studied the effect of the ac drive on the supercurrent. Here we concentrate on the quasi-particles promoted to the continuum via the ionization and refill processes. As shown in the single-channel case~\cite{Riwar2014}, the excess quasiparticles lead to a non-equilibrium charge effect in proximity of the superconducting constriction. Importantly, we here report that this effect is magnified by the number of channels in the multichannel junction.

The production of quasiparticles in the continuum of the right contact~\footnote{For the non-equilibrium quasiparticle production in the left contact, e and h have to be exchanged.} at energy $E$ is
\begin{align}\label{eq_qp_prod}
\dot{P}_{\eta}(E)  = & \Gamma_{\text{I}\eta}\left(T\right)n_A^\text{st}(T)\delta\left[E-\Omega-E_A(T)\right] \\\nonumber
 & +\Gamma_{\text{R}\eta}\left(T\right)\left[2-n_A^\text{st}(T)\right]\delta\left[E-\Omega+E_A(T)\right]\ ,
\end{align}
where $\eta={\text{e},\text{h}}$, and the rates $\Gamma_{\alpha\eta}$ correspond to the partial rates of $\Gamma_{\alpha}$, which distinguish the processes of creating an electron-like ($\eta=\text{e}$) or hole-like ($\eta=\text{h}$) quasiparticle escaping into the right contact. As shown in Ref.~\onlinecite{Riwar2014}, the rates of electron- and hole-like quasiparticle production are different. Due to their corresponding opposite charge $\pm\sqrt{E^2-\Delta^2}/E$, this leads to a finite net charge transfer to the continuum,
\begin{equation}\label{eq_qdot_general}
\dot{q}(T)=\int dE\frac{\sqrt{E^2-\Delta^2}}{E}\left[\dot{P}_\text{e}\left(E\right)-\dot{P}_\text{h}\left(E\right)
\right]\ .
\end{equation}
Here we extend our considerations to the multichannel case. To this end we insert Eq.~\eqref{eq_qp_prod} into Eq.~\eqref{eq_qdot_general}, and sum over all channels to obtain
\begin{figure*}
(a)\includegraphics[width=0.65\columnwidth]{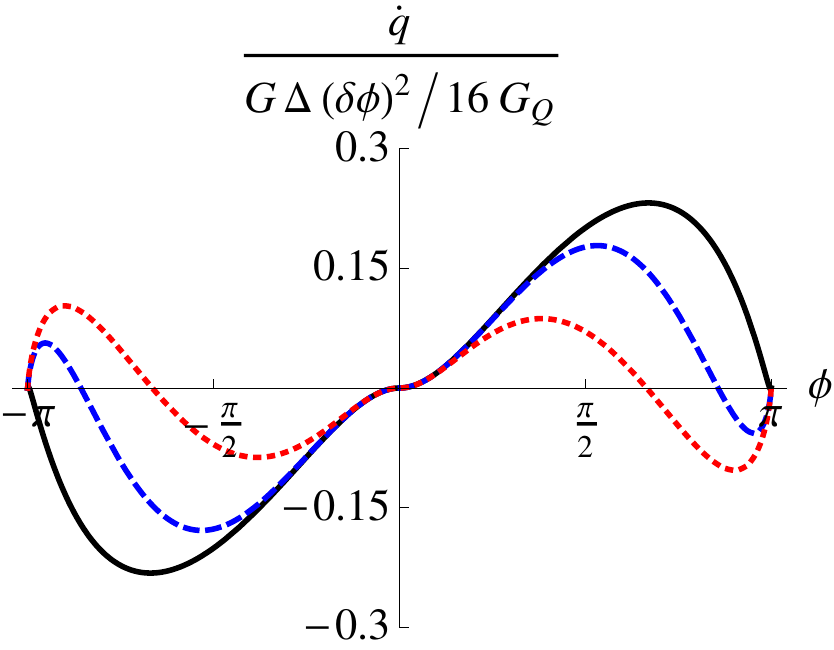}(b)\includegraphics[width=0.65\columnwidth]{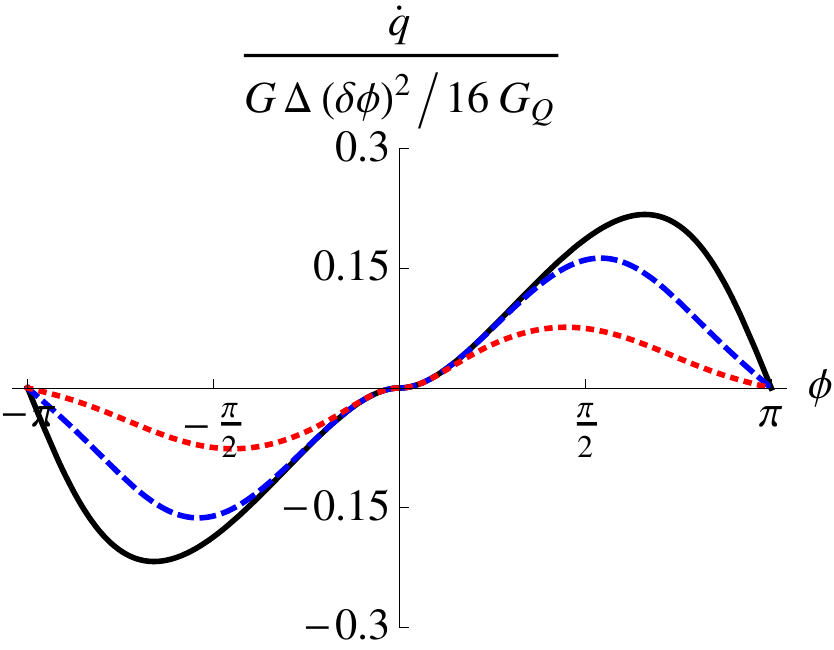}

(c)\includegraphics[width=0.65\columnwidth]{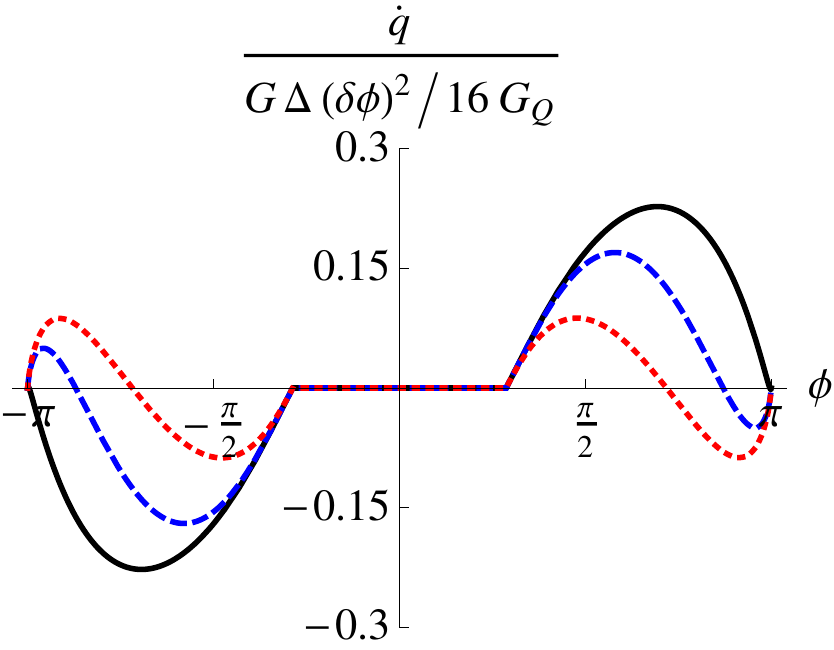}(d)\includegraphics[width=0.65\columnwidth]{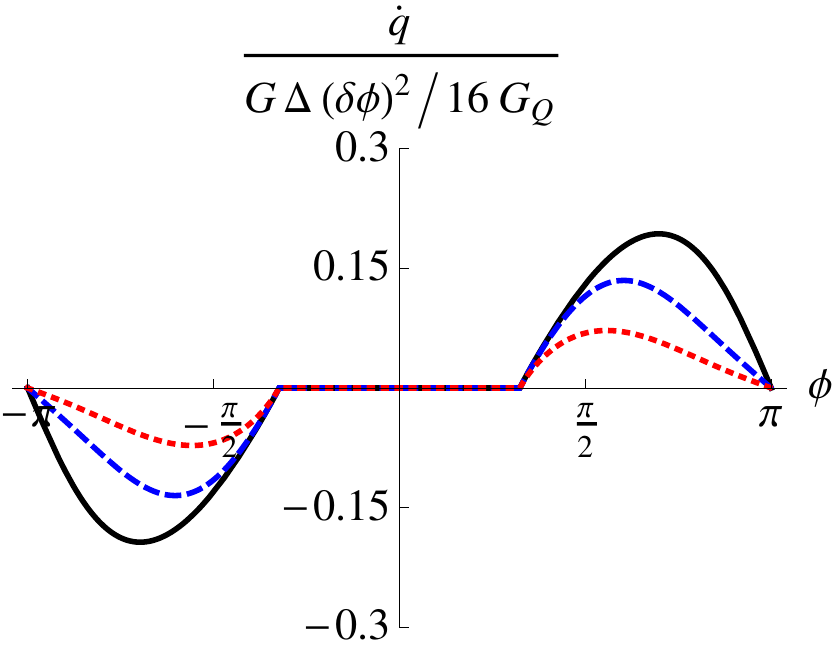}

(e)\includegraphics[width=0.65\columnwidth]{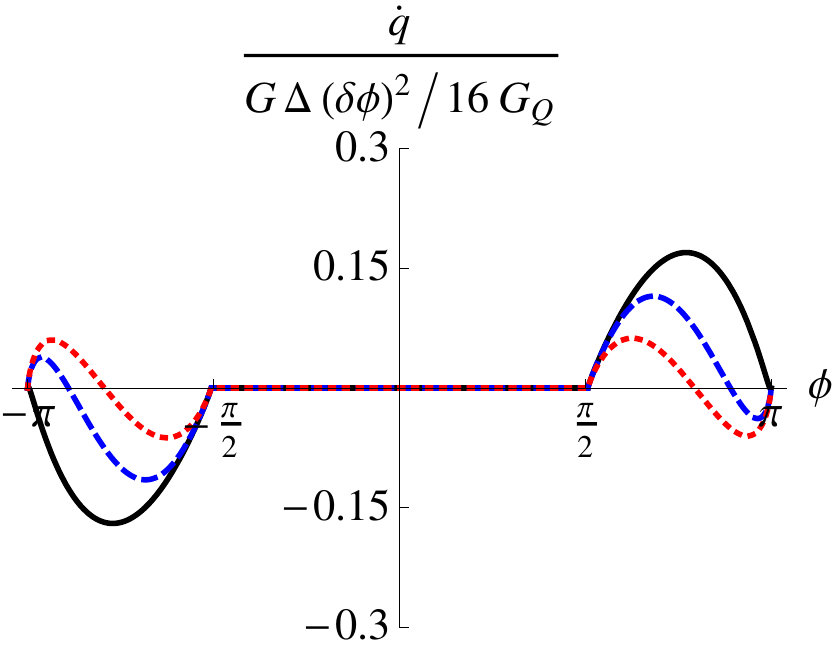}(f)\includegraphics[width=0.65\columnwidth]{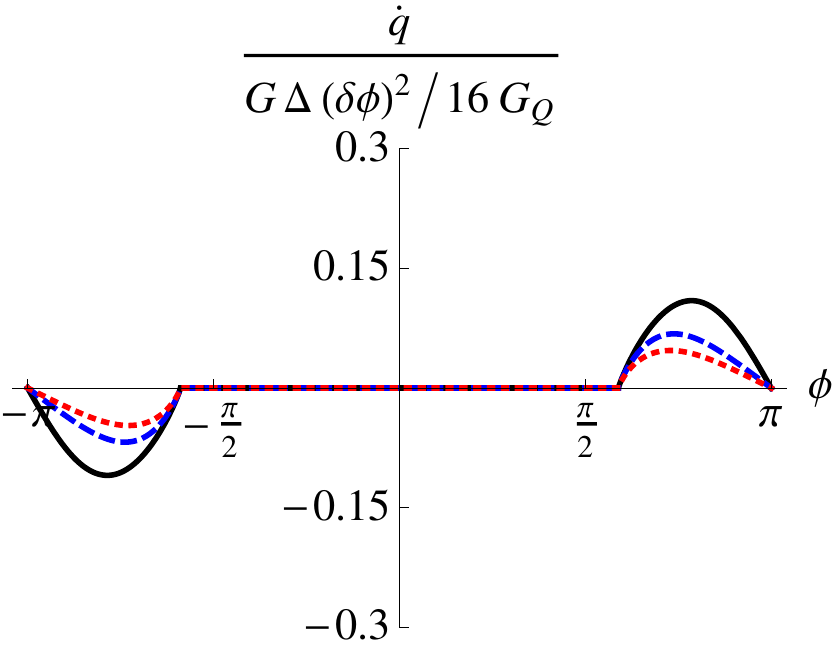}

\caption{The charge transfer $\dot{q}$ as a function of the stationary phase $\phi$ for the diffusive junction (a, c, and e) and for the double junction (b, d, and f). We see that in the diffusive junction the charge transfer can change its sign when refill processes dominate (see main text). The parameters are the same as in Fig.~\ref{fig_Is}.}
\label{fig_qdot}
\end{figure*}
\begin{equation}\label{eq_qdot}
\begin{split}
\dot{q}=\int_{0}^{1}dT\rho\left(T\right)\Biggl(q_{\text{I}}\left(T\right)
\Gamma_{\text{I}}\left(T\right)n_A^{\text{st}}\left(T\right)\\
+q_{\text{R}}\left(T\right)\Gamma_{\text{R}}\left(T\right)\left[2-n_A^{\text{st}}\left(T\right)\right]\Biggr)\ .
\end{split}
\end{equation}
The polarization factors due to the refill and ionization processes are defined as $q_\text{I,R}=\frac{\sqrt{[\Omega\pm E_A]^2-\Delta^2}}{\Omega\pm E_A}(\Gamma_{\alpha\text{e}}-\Gamma_{\alpha\text{h}})/\Gamma_\alpha$, and are given as
\begin{equation}\label{eq_q_I_R}
q_\text{I,R}=
\mp2\frac{\partial E_A}{\partial\phi}\sqrt{\frac{(\Omega\pm E_A)^2-\Delta^2}{\Delta^2-E_A^2}}\frac{E_A\left(1\pm\frac{E_A}{\Omega\pm E_A}\right)}{\Omega E_A\pm\Delta^2(1+\cos\phi)}\ .
\end{equation}
The sign of the ionization contribution $q_\text{I}$ is the same as for the supercurrent carried by the ABS, as indicated with the proportionality factor $\partial E_A/\partial\phi$. On the other hand, the refill contribution $q_\text{R}$ has always an opposite sign for a given phase. Note that only in presence of the refill process there can be a finite $\dot{q}$. The charge transfer is therefore non-zero only above the threshold phase $\phi_0$.

The total charge transfer gives rise to the dissipative current $I_{\text{qp}}=e\dot{q}$,
carried by the free quasiparticles. Note that the net charge scales as $\dot{q}\sim \frac{G}{G_{Q}}\Delta(\delta\phi)^2$, and is therefore in general increased with respect to the single channel charge transfer by the large factor $G/G_{Q}$. On the other hand, the signal is weaker than the ac-induced change of the supercurrent by the factor $(\delta\phi)^2$.

The behaviour of $\dot{q}$ is sensitive to the driving power. When $\Gamma_\text{I,R}\gg\Gamma_\text{A}$, one can easily show through inserting Eqs.~\eqref{eq_stationary_P0_strong} and~\eqref{eq_stationary_P1_strong} into Eq.~\eqref{eq_qdot}, that the contribution of each channel to $\dot{q}$ goes as $\sim q_\text{I}+q_\text{R}$, i.e.,
\begin{equation}
\dot{q}=\int_{0}^{1}dT\rho\left(T\right)\left[q_{\text{I}}\left(T\right)+q_{\text{R}}\left(T\right)\right]\frac{2\Gamma_\text{I}(T)\Gamma_\text{R}(T)}{\Gamma_\text{I}(T)+\Gamma_\text{R}(T)}\ .
\end{equation}
As $|q_\text{I}(T)|\geq|q_\text{R}(T)|$ for all transparencies $T$, $\dot{q}$ has the same sign as $q_\text{I}$. As a consequence, the qualitative behaviour of $\dot{q}$ in this limit does not depend on the details of the junction's channel distribution, see Fig.~\ref{fig_qdot}.

\begin{figure}
\begin{minipage}{\columnwidth}
\vspace{15mm}
\begin{overpic}[width=0.65\columnwidth]%
{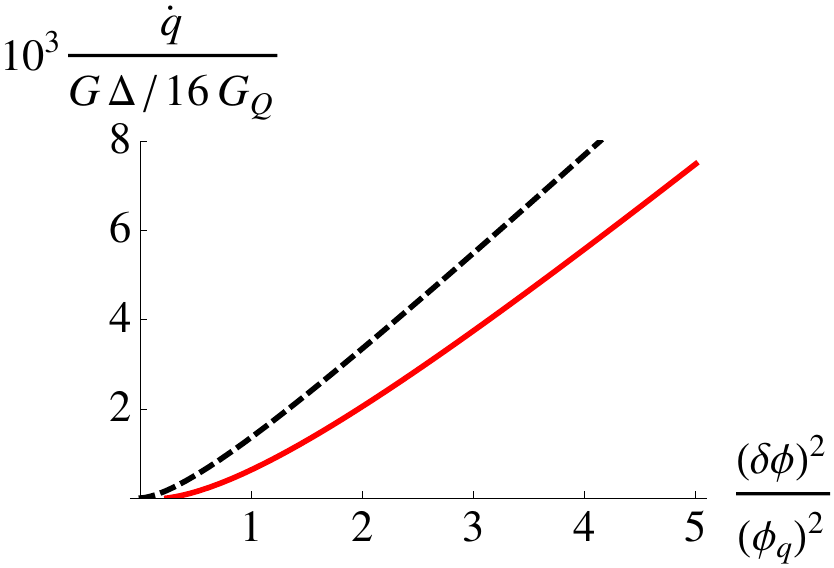}
\put(30,110){\begin{minipage}{1\columnwidth}\fcolorbox{black}{white}{\includegraphics[width=0.45\columnwidth]{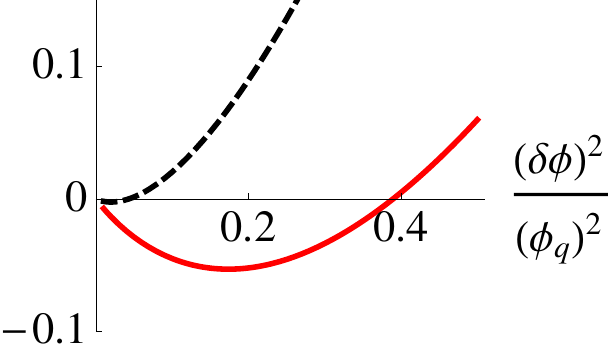}}\end{minipage}}
\end{overpic}
\hspace{15mm}
\end{minipage}

\caption{The charge transfer $\dot{q}$ in a diffusive junction as a function of the driving power $(\delta\phi)^2$. Here, $\phi=8\pi/10$ (solid) and $\phi=7\pi/10$ (dashed). The driving frequency is $\Omega=2\Delta$. The inset is a zoom of the main figure, to show the charge imbalance inversion for weak driving. }
\label{fig_qdot_power}
\end{figure}

In contrast, when the annihilation rate becomes significant (i.e., for weaker driving), we find that the details of the junction become important. When the ac driving power is decreased, the refilling contribution to the charge transfer may exceed the one due to ionization, which is simply due to the fact that a significant phase noise suppresses the double occupation. We readily see this in Eq.~\eqref{eq_qdot}, where the refill contribution may exceed the ionization one when $n_A<1$, which is the case for $\Gamma_\text{I,R}\ll\Gamma_\text{A}$, see also Eqs.~\eqref{eq_stationary_P0_weak} and~\eqref{eq_stationary_P1_weak}. Note, however, that the charge imbalance inversion only occurs in the channels with a sufficiently high transparency~\footnote{The lowest transparency for an individual channel at which the charge imbalance inversion occurs is $T\approx 0.725$ for $\Omega=2\Delta$. This value increases for lower driving frequencies $\Omega<2\Delta$.}. For low $T$, the quasiparticles due to refill are always created near the band edge, where the charge asymmetry of the electron- and hole-like quasiparticles is suppressed, $\sqrt{E^2-\Delta^2}/E\rightarrow0$, and consequently, $q_\text{R}\rightarrow 0$. Based on this it becomes apparent for the multichannel case, that only when the junction contains a significant number of high $T$ channels, this effect will survive. This is the case for the diffusive wire, see Figs.~\ref{fig_qdot}a, c, and e, where we observe that the charge imbalance inversion persists. In Fig.~\ref{fig_qdot_power} we explicitly show the onset of this charge imbalance reversal when lowering the driving power (for the stationary phase $\phi=4\pi/5$, solid line). Note that the inversion does not take place for all stationary phases, as is shown in the black dashed curve for $\phi=3\pi/5$.
For the double junction with a maximal transparency $T_\text{max}<1$, no charge imbalance inversion occurs in general, see Figs.~\ref{fig_qdot}b, d, and f (the exception is a nearly symmetric junction with $T_c\approx 1$). We hence see a rather distinct qualitative difference for different junctions.

\section{Time evolution of supercurrent after ac pulse}\label{sec_supercurrent_time}

Motivated by recent experimental progress in supercurrent spectroscopy both in single~\cite{Bretheau2013b}- and multichannel junctions~\cite{LevensonFalk2014}, we study the time evolution of the supercurrent, stressing the longevity of the odd ABS occupation. For that purpose we consider the following driving protocol. First, the system is driven with an ac power $\delta\phi\gg\phi_{q}$ such that $\Gamma_\text{I,R}$ exceeds $\Gamma_\text{A}$, to initialise a non-zero double occupation. The
occupation probabilities for each channel in this limit are given by Eqs.~\eqref{eq_stationary_P0_strong} and~\eqref{eq_stationary_P1_strong}. 
This stationary distribution is achieved for a driving pulse duration exceeding the time scale $\max\left(\Gamma_\text{I}^{-1},\Gamma_\text{R}^{-1}\right)$, establishing the stationary supercurrent $I_S^\text{st}$.

Subsequently, we assume that the driving is switched off, and we look at the relaxation of the supercurrent on a time scale shorter than the parity relaxation time, such that singly occupied states cannot relax. On this time scale the double occupation of the ABS in each channel decays exponentially due to the phase noise induced annihilation process, $\Gamma_\text{A}(T)$, and the time resolved current can be given as
\begin{align}
I_{S}\left(t\right)=&-2e\int_0^{1}dT\rho(T)\frac{\partial E_{A}\left(T\right)}{\partial\phi}\\\nonumber & \times\left[1-P_{1}^{\text{st}}\left(T\right)-2e^{-\Gamma_{\text{A}}\left(T\right)t}P_{2}^{\text{st}}\left(T\right)\right]\ ,
\end{align}
where $t=0$ signifies the time when the ac drive is switched off. In the long time limit $I_S^\text{long}=I_S(t\gg\Gamma_\text{A}^{-1})$ the current relaxes to a stationary value different
from the equilibrium value,
\begin{equation}\label{eq_I_S_long}
I_{S}^\text{long}=-2e\int_0^{1}dT\rho(T)\frac{\partial E_{A}\left(T\right)}{\partial\phi}\left[1-P_{1}^{\text{st}}\left(T\right)\right]\ ,
\end{equation}
due to $P_1^\text{st}\neq0$. The supercurrent approaches the stationary value $I_{S}^\text{long}$ which is below the equilibrium current.

\begin{figure}
\includegraphics[width=0.65\columnwidth]{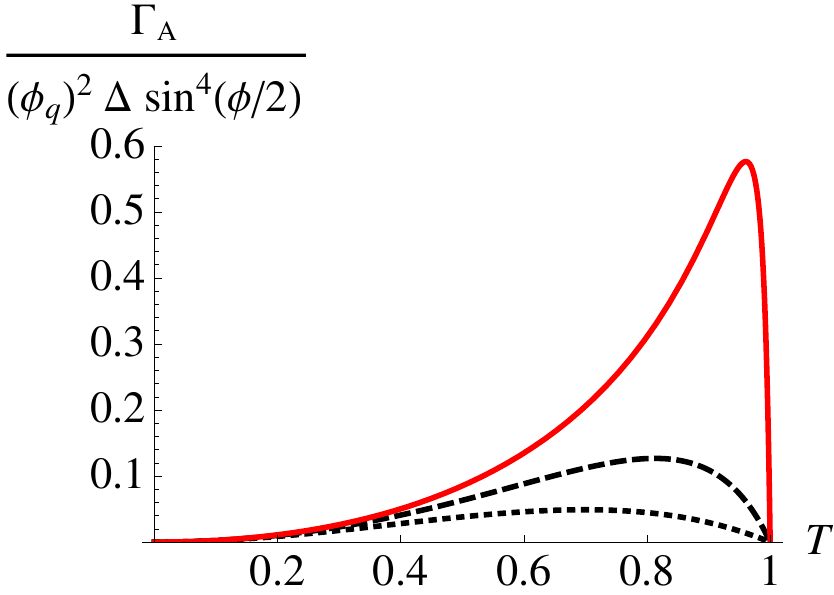}
\caption{The annihilation rate $\Gamma_\text{A}$ as a function of $T$ for the stationary phases $\phi=\pi/3,2\pi/3,0.9\pi$ (dotted, dashed, solid). Note that in the figure the rate is divided by $\sin^4(\phi/2)$. As the rate itself scales with this factor, its values for phases $\phi$ significantly different from $\pi$ are suppressed.}
\label{fig_Gamma_A}
\end{figure}

In order to facilitate the discussion of the time evolution of the supercurrent, we consider first the annihilation rate $\Gamma_\text{A}$ as a function of the transparency (see Fig.~\ref{fig_Gamma_A}), which defines the behaviour of $I_S$. Note that the annihilation rate vanishes in both the limit of small transparency and for fully transparent channels. In the limit of $T$ close to zero, $\Gamma_\text{A}\sim T^2$, while for $T$ close to $1$, $\Gamma_\text{A}\sim 1-T$. There is a global maximum for a finite, intermediate transparency. The rate in total has a magnitude that depends strongly on the stationary phase bias $\phi$. Namely, for decreasing $\phi$, the rate likewise decreases globally as $\sim\phi^4$.

\begin{figure*}
(a)\includegraphics[width=0.65\columnwidth]{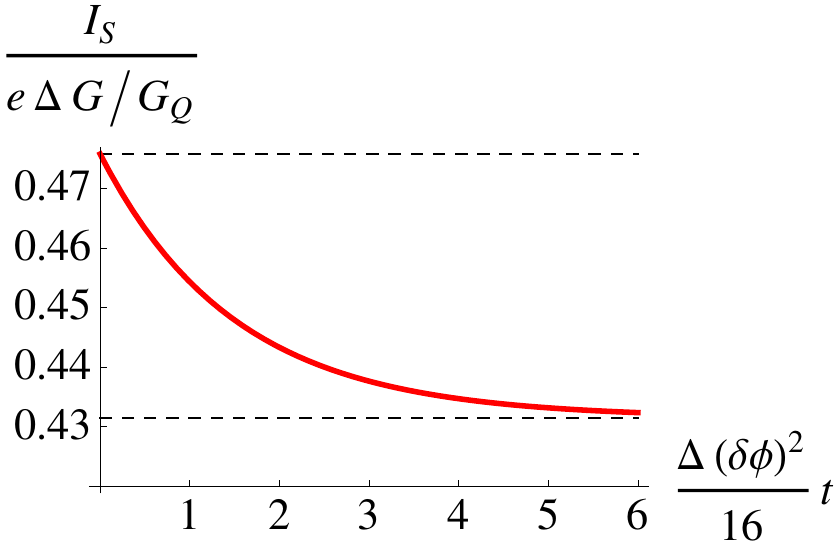}(b)\includegraphics[width=0.65\columnwidth]{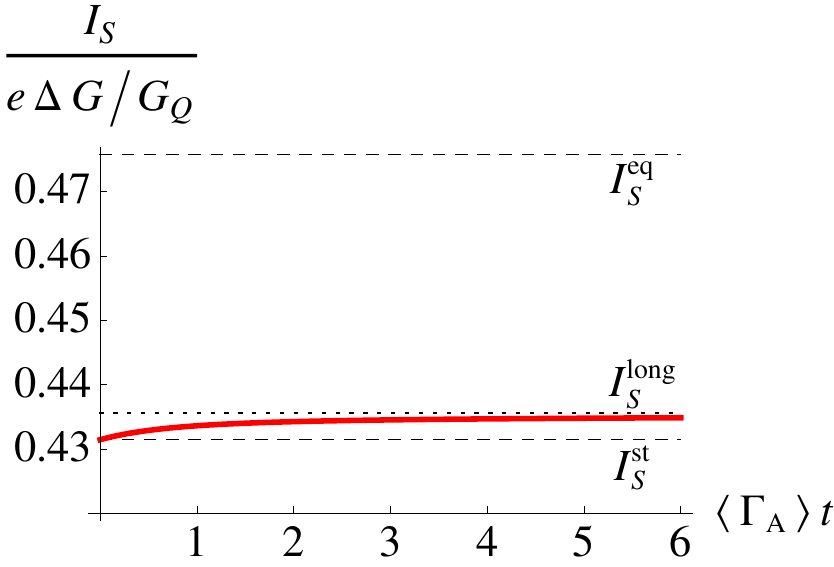}

(c)\includegraphics[width=0.65\columnwidth]{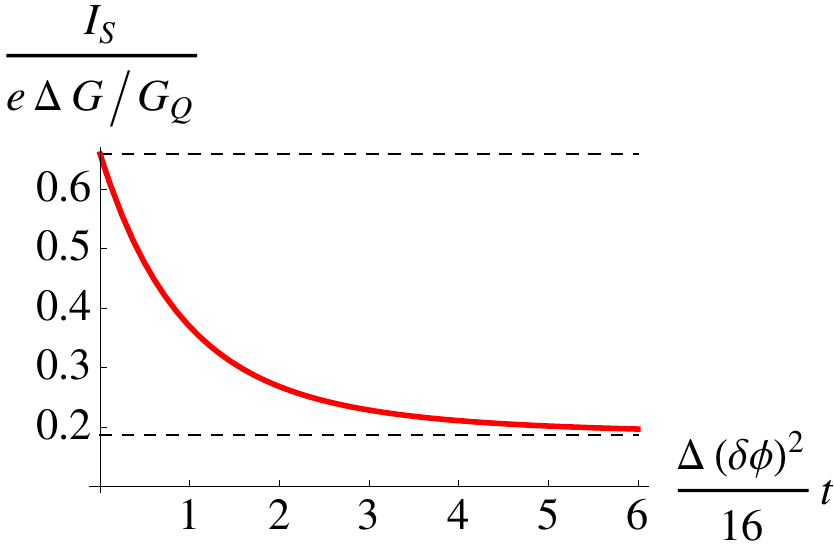}(d)\includegraphics[width=0.65\columnwidth]{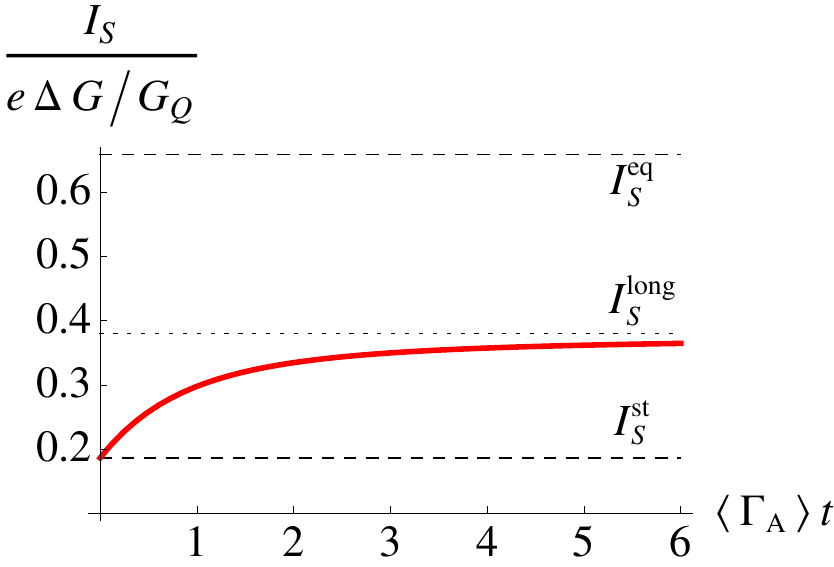}

\caption{The time evolution of the supercurrent for the diffusive junction. Figures a and c show the time evolution during the ac pulse, where the supercurrent departs from the equilibrium value $I_S^\text{eq}$ and eventually reaches the stationary value for the driven system $I_S^\text{st}$ on a time scale $1/[\Delta(\delta\phi)^{2}]$. The equilibrium current $I_S^\text{eq}$ and the stationary driven current $I_S^\text{st}$ are shown as the upper/lower dashed line, respectively. The driving frequency is $\Omega=1.9\Delta$. Figures b and d show the time evolution of the supercurrent once the pulse is switched off, on the time scale of the decay of the doubly occupied ABS. In figures a and b (c and d), the stationary phase bias is $\phi=\pi/3$ ($2\pi/3$). The supercurrent decays with the respective rate $\langle\Gamma_\text{A}\rangle=1.22\times10^{-3}\Delta(\phi_q)^2$ (b) and $\langle\Gamma_\text{A}\rangle=4.20\times10^{-2}\Delta(\phi_q)^2$ (d). As the singly occupied ABS does not decay on this time scale, the value of the supercurrent decays to the new stationary value $I_S^\text{long}$ in the long time limit, see Eq.~\eqref{eq_I_S_long}, which is indicated as the dotted line. }
\label{fig_Is_time}
\end{figure*}

\begin{figure*}

(a)\includegraphics[width=0.65\columnwidth]{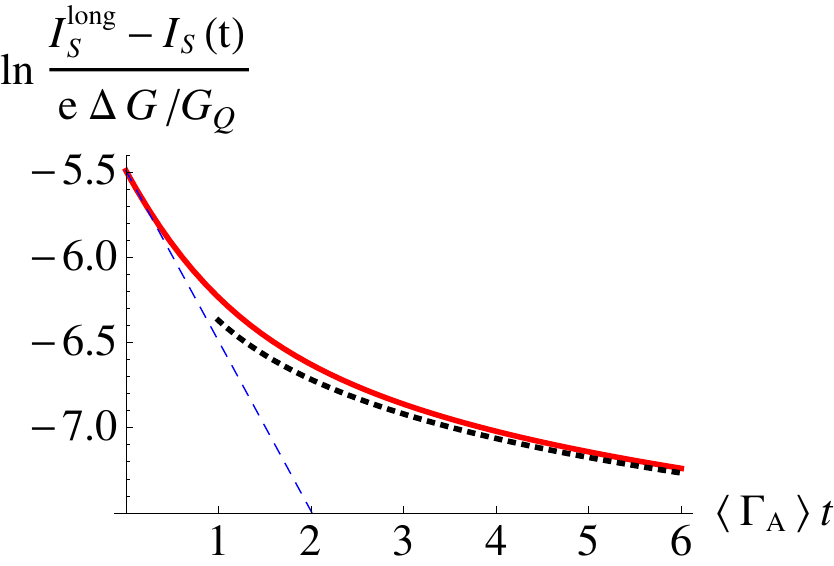}
(b)\includegraphics[width=0.65\columnwidth]{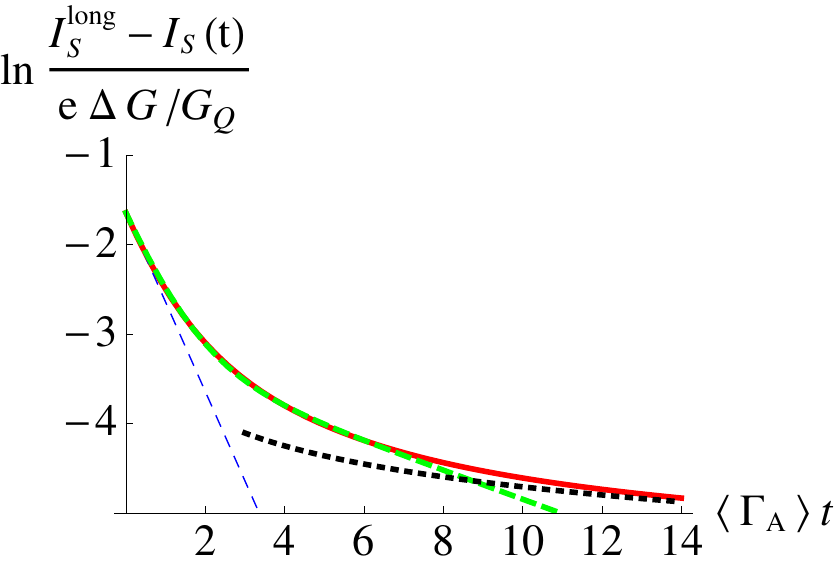}

\caption{The time evolution of the supercurrent (Figs.~\ref{fig_Is_time}b and~d) after switching off the ac pulse on a logarithmic scale (solid line). For a stationary phase $\phi=\pi/3$ (a), the time evolution can be described by an exponential decay for short times (thin dashed line) and an algebraic $1/\sqrt{t}$ decay for long times (dotted line). For $\phi=2\pi/3$ (b), there is an additional time scale appearing due to the decay of the low $T$ channels, captured by the two-exponential fit (thick dashed line) with $I_S^\text{long}-I_S(t)=a_\text{fast} e^{-\Gamma_\text{fast} t}+a_\text{slow} e^{-\Gamma_\text{slow} t}$
where $\Gamma_\text{fast}\approx 1.275\langle\Gamma_A\rangle$ and $\Gamma_\text{slow}\approx 0.205\langle\Gamma_A\rangle$.}
\label{fig_Is_time_log}
\end{figure*}

Based on the properties of $\Gamma_\text{A}$, there is a distinct difference of the short and long time behaviour in the time evolution, as we will explain in the following (see also Fig.~\ref{fig_Is_time}). Shortly after switching off the drive, the supercurrent relaxes exponentially as expected, and the time evolution can be approximated as
\begin{equation}
I_S(t)\approx I_S^\text{long}+e^{-\langle \Gamma_\text{A}\rangle t}\left[I_S^\text{st}-I_S^\text{long}\right]\ .
\end{equation}
The effective relaxation rate is given as the average
\begin{equation}
\langle \Gamma_\text{A}\rangle=\frac{\int_0^{1} dT\,\widetilde{\rho}(T) \Gamma_\text{A}(T)}{\int_0^{1} dT\,\widetilde{\rho}(T)}\ ,
\end{equation}
with the effective distribution function $\widetilde{\rho}(T)=\left|\frac{\partial E_\text{A}}{\partial\phi}(T)\right|P_2^\text{st}(T)\rho(T)$. The effective distribution signifies that \textit{all} channels contribute to the initial decay of the supercurrent, weighted by their initial occupation $P_2^\text{st}$ and the magnitude of the supercurrent they support, $\left|\frac{\partial E_\text{A}}{\partial\phi}\right|$. 
This initial behaviour is indicated in the log plot of Fig.~\ref{fig_Is_time_log}a and~b, where we see that it is a good approximation for $t\leq\langle\Gamma_\text{A}\rangle^{-1}$. 

Interestingly, the long time behaviour for $t>\langle\Gamma_\text{A}\rangle^{-1}$ in a diffusive junction is radically different. Namely, the long time relaxation goes strikingly as $\sim 1/\sqrt{t}$, as we will outline just now. The origin for this surprising feature lies in the long lifetime of the double occupation of the channels with $T$ close to $1$, where, as mentioned before, $\Gamma_\text{A}(T)\sim1-T$. Consequently, the current contributions of the lower transparency channels decay one after the other, such that with increasing time the phase noise gradually selects a population with increasing $T$ which in turn relaxes slower. This long time selection of highly transparent channels effectively leads to an algebraic decay of the current. In particular, we find
\begin{equation}
I_{S}\left(t\right)\approx I_{S}^\text{long}+\left.e\frac{G}{G_{Q}}\frac{\partial E_A}{\partial\phi}(T)\frac{\sqrt{\pi}P_2^\text{st}(T)}{\sqrt{\frac{\partial\Gamma_\text{A}}{\partial T}(T) }}\frac{1}{\sqrt{t}}\right|_{T\rightarrow1}\ .
\end{equation}
If the initial occupation of bound states involves only channels with sufficiently high transparency, the above two asymptotics are enough to characterize the entire time evolution. 
However, as we mentioned before, $\Gamma_\text{A}(T)$ decreases also for low transparency channels. Consequently, if the low tranparency channels are occupied initially (which is the case at a sufficiently high $\Omega$ and for a sufficiently nonzero $\phi$), they introduce a second exponential decay on a time scale between the above two asymptotics. That is, their decay is slower than the initial decay, but can of course not outlast the algebraic decay of the high $T$ channels for longer times. This can be seen in Fig.~\ref{fig_Is_time_log}.

In the example of a stationary phase $\phi=\pi/3$ in Fig.~\ref{fig_Is_time_log}a, the two asymptotics (exponential for initial, and algebraic for long time evolution) are sufficient to capture the dynamics of the supercurrent. For $\phi=2\pi/3$, in Fig.~\ref{fig_Is_time_log}b we detect a regime of intermediate times where neither asymptotics apply. In the example of  Fig.~\ref{fig_Is_time_log}b, the initial occupation of bound states includes channels with $T>0.25$, where the slower relaxation emerges from these low $T$ channels. The two different time scales can be taken into account by decomposing the time evolution into a decay of the channels with a shorter lifetime, and a second slow decay of the low $T$ channels. This fact is accounted for by the numeric fit, which comprises of the sum of two exponentials, $I_S(t)=I_S^\text{long}-a_\text{fast} e^{-\Gamma_\text{fast} t}-a_\text{slow} e^{-\Gamma_\text{slow} t}$, with a fast ($\Gamma_\text{fast}>\langle\Gamma_\text{A}\rangle$) and a slow decay related to the low transparency channels ($\Gamma_\text{slow}<\langle\Gamma_\text{A}\rangle$)~\footnote{The fit is subject to the constraints $a_\text{fast}+a_\text{slow}=I_S^\text{long}-I_S^\text{st}$ and $a_\text{fast}\Gamma_\text{fast}+a_\text{slow}\Gamma_\text{slow}=(a_\text{fast}+a_\text{slow})\langle\Gamma_\text{A}\rangle$, to capture the initial decay exactly.}. Note that in general, the fast fraction of the exponential decay is larger than the slow one, $a_\text{fast}>a_\text{slow}$, because the slowly decaying low $T$ channels support a weaker supercurrent.

Note that, for an even higher stationary phase $\phi\rightarrow \pi$, the algebraic decay is suppressed. This can be understood when taking the annihilation rate in the limit of $T$ close to 1 and $\phi$ close to $\pi$, $\Gamma_{\text{A}}\approx\frac{\Delta\left(\phi_{q}\right)^{2}}{8}\left(1-T+(\pi-\phi)^{2}/4\right)^{-3/2}\left(1-T\right)$. We thus see that the interval of transparencies with a very long lifetime is growing smaller and smaller for $\phi$ close to $\pi$, because the rate increases quickly in the neighbourhood of $T=1$, namely $\Gamma_\text{A}\sim \left(\pi-\phi\right)^{-3}(1-T)$ for $1\gg \left(\pi-\phi\right)^{2}\gg 1-T$ (see solid curve in Fig.~\ref{fig_Gamma_A}). This leads to a suppression of the algebraic decay by the factor $\sim \left(\pi-\phi\right)^{3/2}$. In this case, the only long time behaviour is due to the low transparency channels.

\begin{figure}
(a)\includegraphics[width=0.65\columnwidth]{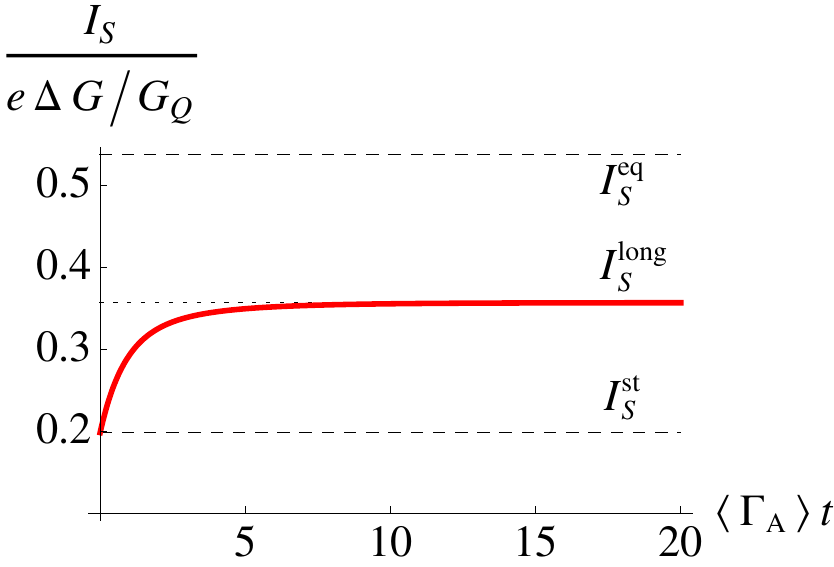}

(b)\includegraphics[width=0.65\columnwidth]{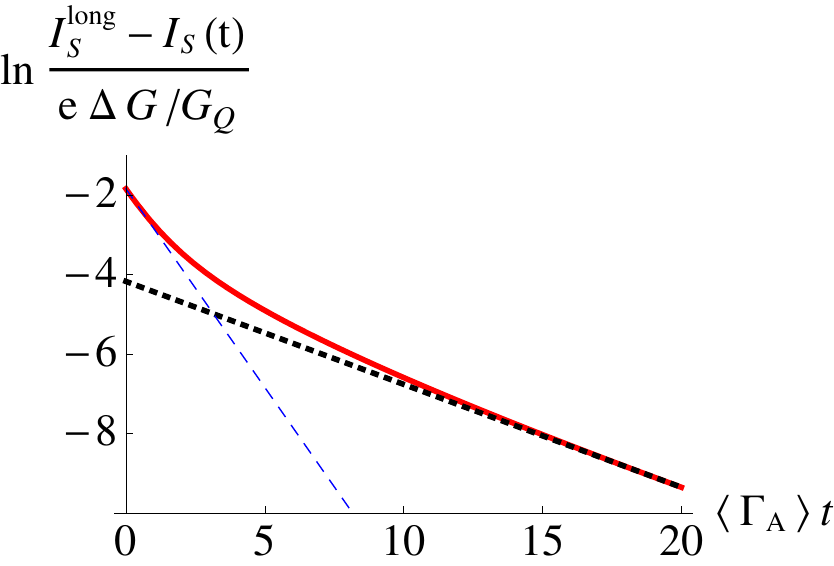}

\caption{The time evolution of the
supercurrent for the double junction, after the ac pulse ($\Omega=1.9\Delta$) in a linear scale (a) and a logarithmic scale (b). In (b), both the long (dotted line) and short term (dashed line) asymptotics are exponential decays. The stationary phase bias is $\phi=2\pi/3$. }
\label{fig_Is_double_time}
\end{figure}

In the case of the double junction on the other hand, there is no algebraic decay at any $\phi$, as there are no highly transparent channels. Here, the long time evolution is exclusively dominated by the low transparency channels, which gives rise to an ordinary exponential time dependence. This fact is visualised in Fig.~\ref{fig_Is_double_time}.

\section{Effect of parity relaxation}\label{sec_parity}
In the above sections, we neglected parity relaxation. Thus, singly occupied ABS could not relax, which prevented the system from thermalizing. The justification was that we expect the parity relaxation rate $\Gamma_\text{P}\ll\Gamma_\text{A}$, as argued in the introduction. Now we want to qualitatively describe the effect of a finite $\Gamma_\text{P}$ for the ultraweak driving regime $\Gamma_\text{I,R}\sim\Gamma_\text{P}\ll\Gamma_\text{A}$. 

\begin{figure}
\includegraphics[width=0.65\columnwidth]{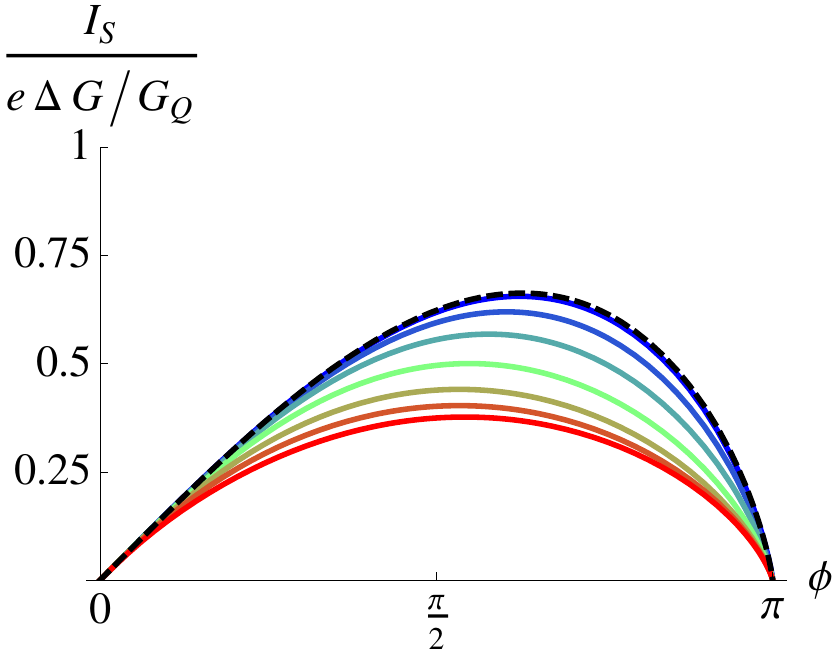}
\caption{The supercurrent $I_S$ as a function of the phase. This figure shows the transition between the weak driving (red) to ultraweak driving (blue) limit. From top to bottom the curves are shown for $(\delta\phi)^2=\{\frac{1}{64},\frac{1}{9},\frac{1}{3},1,3,9,64\}\times16\frac{\Gamma_\text{P}}{\Delta}$. The driving frequency is $\Omega=2\Delta$. As the driving power decreases, the current approaches its equilibrium value (dashed line).}
\label{fig_Is_thermal}
\end{figure}

We do so by adding the corresponding process in the rate equation on a phenomenological level,
\begin{equation}
\dot{P}_{1}\left(T\right) = 2\Gamma_{\text{R}}\left(T\right)P_{0}\left(T\right)-\left[\Gamma_{\text{R}}\left(T\right)+\Gamma_{\text{I}}\left(T\right)+\Gamma_\text{P}\right]
P_{1}\left(T\right)\ ,
\end{equation}
and $P_0+P_1=1$, while $P_2=0$.
Since the origins of $\Gamma_\text{P}$ can be manifold (see introduction), we keep the considerations at the simplest level where we do not impose any dependence of $\Gamma_\text{P}$ on $T$ or $\phi$.
The stationary solutions are
\begin{equation}
P_{1}^{\text{st}}\left(T\right) =\frac{2\Gamma_{\text{R}}}{3\Gamma_\text{R}(T)+\Gamma_\text{I}(T)+\Gamma_\text{P}}\ ,
\end{equation}
and $P_0^\text{st}=1-P_1^\text{st}$.
From this we realize immediately that the bound state distribution thermalizes for $\Gamma_\text{I,R}\ll\Gamma_\text{P}$, when $P_1$ is suppressed. The consequence for $I_S$ is visualised in Fig.~\ref{fig_Is_thermal}, where, for decreasing driving power, the supercurrent returns to its equilibrium value. 

A rigorous discussion of the charge imbalance in the ultraweak driving regime is not possible. The reason for this can be argued as follows. The origin of the charge imbalance is explicitly the charge asymmetry in the parity breaking processes due to the driving ($\Gamma_\text{I,R}$), and the fact that the ABS and continuum are coupled. In general, the parity relaxation processes may involve the continuum as well. In that case, they may contribute to the charge imbalance. If however the dominant processes for parity relaxation do not involve the continuum, for instance, if the leading processes are due to the interchannel coupling that relaxes the ABS occupation, then Eq.~\eqref{eq_qdot} remains valid. In that case, $\Gamma_\text{P}$ simply alters the occupation of the bound states. The parity relaxation then leads to a lowering of $P_1$, making the ionization process ineffective. Consequently, the contribution to $\dot{q}$ due to the refill process gains in importance. Here, it can therefore be expected that the previously described sign change of the charge transfer (see Sec.~\ref{sec_charge_imbalance}) is actually more likely to occur. In the most extreme case, $\Gamma_\text{I,R}\ll\Gamma_\text{P}$, the only remaining contribution to the charge transfer is due to the refill process,
\begin{equation}
\dot{q}=2\int_{0}^{1}dT\rho\left(T\right)q_{\text{R}}\left(T\right)\Gamma_{\text{R}}\left(T\right)\ ,
\end{equation}
and thus has the opposite sign as the equilibrium supercurrent for all values of the phase bias.

\section{Conclusions}\label{sec_conclusions}

We studied a multichannel weak link subject to an ac driven phase bias. We predict a weak driving regime, where there is a strong change of the supercurrent due to the ac drive, of the same order of magnitude as the equilibrium supercurrent. Moreover, when the ac-induced processes occur with a faster rate than the environment induced annihilation processes, we find that the total supercurrent of the ac driven system can be reversed with respect to the equilibrium system. In addition, for a driving frequency close to twice the gap, the supercurrent reveals a specific power law with respect to the frequency that is directly related to the distribution of the low transparency channels, and thus provides information about the nature of the junction.

On top of the supercurrent we show that there is a finite quasiparticle charge transfer due to the asymmetry of the ionization and refill processes, which leads to a charge imbalance carried by the continuum states. Importantly, the magnitude of the charge imbalance signal is highly magnified with respect to the single channel case~\cite{Riwar2014} by the large number of channels. We find a sign inversion of the charge transfer, emerging only if there are channels close to $T=1$, a feature that therefore provides information about whether the junction has highly transparent channels.

We finally consider the time evolution of the supercurrent after applying an ac pulse, on the time scale of the lifetime of the doubly occupied bound states, where, importantly, the current does not return to the equilibrium value due to the much slower parity relaxation. While the onset of the time evolution is exponential, the asymptotic behaviour exhibits the power law $1/\sqrt{t}$ due to a long-time survival of the doubly occupied states of the channels with a transparency close to 1. This effect vanishes for junctions without transparent channels, where the long time relaxation is again exponential.

Finally, we comment on the effects of parity relaxation, which enables the thermalization of the bound state population when the rate of the ac processes becomes comparable to the parity relaxation. Consequently, in this regime, the ac-induced change in the stationary supercurrent becomes small. We propose possible mechanisms that could lead to parity relaxation, such as recombination with free quasiparticles in the contacts, or recombination with or transitions between bound states of different channels. If the latter are the dominant processes, we expect that the charge imbalance effect remains. It still scales with the number of channels, with the difference that here, the charge imbalance due to the refill processes become dominant. An open question for further studies is the validity of our predictions in the long-junction limit.

This work has been supported by the Nanosciences Foundation in Grenoble, in the frame of its Chair of Excellence program.

\appendix
\section{Cusp in the phase dependence of the supercurrent and the charge transfer near $\phi_0$}\label{app_cusp}
In this appendix, we provide the derivation of the cusp feature of the supercurrent and the charge transfer in the irradiated junction at $\phi=\phi_0$, as discussed in Secs.~\ref{sec_supercurrent} and~\ref{sec_charge_imbalance}. We will show that the cusp feature is a consequence of the diverging density of the high $T$ channels in either the diffusive or double junctions. We focus in the following on the cusp for positive $\phi$ (due to the anti-symmetry of both $I_S$ and $\dot{q}$ with respect to $\phi$, the cusp for negative $\phi$ follows in analogy).

Let us first provide the explanation for the cusp feature of the supercurrent. For this purpose, consider the general expression for the supercurrent in the stationary case, $n_A=n_A^\text{st}$,
\begin{equation}\label{eq_I_S_st}
I_S^\text{st}=-2e\int_{0}^{1}dT\rho\left(T\right)\frac{\partial E_{A}}{\partial\phi}\left(T\right)\left[1-n_A^\text{st}(T)\right]\ .
\end{equation}
An inspection of above equation for $\phi$ close to $\phi_0$ reveals the following. 
Firstly, close to the threshold only the high transparency channels get refilled. The finite interval of transparencies close to the maximal $T=T_\text{max}$, which contribute to the supercurrent, scales linearly. That is, it can be shown that the lower integration limit is $T_\text{max}[1-\alpha(\phi-\phi_0)]$, where $\alpha=\sqrt{\frac{T_\text{max}\Delta^2}{\Omega(2\Delta-\Omega)}-1}$. Note that this approximation is valid for $\alpha(\phi-\phi_0)\ll 1$, which means that $\Omega$ has to be sufficiently below $2\Delta$. 
Therefore, we need to evaluate the behaviour of $n_A$ with respect to $\phi$ at the threshold $\phi_0$ for the corresponding maximal transparency. Note that $n_A$ may be evaluated at $T=T_\text{max}$ due to the small integration interval. It can be shown that $n_A\sim\sqrt{\phi-\phi_0}$ for $\phi>\phi_0$ (and $n_A=0$ for $\phi\leq\phi_0$). Close to the threshold $\phi_0$, the refill rate likewise vanishes as $\Gamma_\text{R}\sim\sqrt{\phi-\phi_0}$ [which follows from Eq.~\eqref{eq_Gamma_I_R}], while $\Gamma_\text{I}$ remains finite. Consequently,  the occupation number can be expanded in the small rate $\Gamma_\text{R}$, which leads to $n_A\sim\Gamma_\text{R}$. 

Taking into account both the approximation for the integration limits and $n_A$, the change of the supercurrent from its equilibrium value $I_S^\text{eq}\equiv I_S(n_A=0)$ may be written close to the threshold as 
\begin{align}
I_S-I_S^\text{eq}&\sim \sqrt{\phi-\phi_0}\int_{T_\text{max}[1-\alpha(\phi-\phi_0)]}^{T_\text{max}} dT \frac{1}{\sqrt{T_\text{max}-T}}\\\nonumber &\sim \phi-\phi_0\ ,
\end{align}
for $\phi\geq\phi_0$ (while $I_S-I_S^\text{eq}=0$ for $\phi<\phi_0$). Note that we approximated the channel transparency distribution for $T$ close to $T_\text{max}$ which provides the term $\sim 1/\sqrt{T_\text{max}-T}$ for \textit{both} the diffusive and the double junction. Thus we see that the cusp feature is a direct consequence of the square root divergence in $\rho(T)$ at the maximal $T$.

One may in analogy argue that the onset of $\dot{q}$ occurs as the same cusp feature. In order to do so, we consider Eq.~\eqref{eq_qdot}. The argumentation of the integration limits is the same as above. However, we see that the integrand contains more terms that depend on the driving frequency, and thus may contain a threshold. Namely, we see that on top of $n_A$ and $\Gamma_\text{R}$ we need to discuss likewise $q_\text{R}$ close to the threshold at $\phi_0$. Note that the terms $\Gamma_\text{I}$ and $q_\text{I}$ may be evaluated directly at $\phi=\phi_0$. The corrections of these terms close to the threshold scale linearly $\sim\phi-\phi_0$, which means that they can be discarded as higher order corrections. As for the polarization factor of the refill process, $q_\text{R}$, one may convince oneself through Eq.~\eqref{eq_q_I_R}, that also $q_\text{R}\sim\sqrt{\phi-\phi_0}$ for $\phi>\phi_0$. In total, when comparing all contributions, we see that the lowest order term that survives in the integrand of Eq.~\eqref{eq_qdot} is $q_\text{I}\Gamma_\text{I}n_A\sim\sqrt{\phi-\phi_0}$. The term related to the refill process scales as $\sim\phi-\phi_0$ (due to the product of $\Gamma_\text{R}$ and $q_\text{R}$) and is therefore discarded. As a consequence, we find the same scaling behaviour of the integral limits and the integrand as for the supercurrent discussed above. Therefore, the same cusp behaviour emerges, namely $\dot{q}\sim\phi-\phi_0$ for $\phi>\phi$ and $\dot{q}=0$ for $\phi\leq\phi_0$. Note also, that the leading order term being proportional to $q_\text{I}$ means that close to the threshold $\phi_0$, the dominant contribution to the charge imbalance is always due the ionization process.

\section{Supercurrent at driving frequencies close to $2\Delta$}\label{app_I_S_high_frequency}
In Eqs.~\eqref{eq_Is_Omega_rho_D} and~\eqref{eq_Is_Omega_rho_dj} of Sec.~\ref{sec_supercurrent}, we provide analytic expressions for the supercurrent in the driven junction at driving frequencies close to, but smaller than $2\Delta$. Here we outline the derivation of these expressions. The starting point is again the equation for the stationary supercurrent, Eq.~\eqref{eq_I_S_st}.
It is straightforward to see that only the term $n_A$ depends on $\Omega$. Now, as we pointed out already in the main text, for a driving frequency close to $\Omega\sim2\Delta$, there is a non-zero occupation of the bound states, $n_A^\text{st}\neq0$, almost up to channels with zero transparency. Therefore, in the regime of $\Omega\sim2\Delta$, the supercurrent may be approximated as the value for $\Omega=2\Delta$ and a small correction that takes into account the low $T$ channels only. The lowest order correction in $\delta\Omega=2\Delta-\Omega$ can be written as,
\begin{equation}
I_S\approx \left.I_S\right|_{\Omega=2\Delta}+2e\int_0^{\delta T}dT \rho(T)\frac{\partial E_A}{\partial\phi}(T) \left[\left.n_A^\text{st}(T)\right|_{\Omega=2\Delta}\right]\ .
\end{equation}
Note that the upper integration limit $\delta T=\frac{2\delta\Omega}{\Delta \sin(\phi/2)^2}$ is already first order in $\delta\Omega$, which justifies to take the zeroth order term of $n_A$ in $\delta\Omega$, that is, $n_A$ for $\Omega=2\Delta$ in the integrand. Likewise, as $\delta T\ll 1$, we may expand the integrand to the lowest order in $T$. For the diffusive junction, we thus obtain
\begin{equation}
I_S\approx \left.I_S\right|_{\Omega=2\Delta}+\frac{eG}{G_{Q}}\frac{\partial^2 E_A}{\partial\phi\partial T}(0) \left.n_A(0)^\text{st}\right|_{\Omega=2\Delta}\int_0^{\delta T}dT\ .
\end{equation}
The above equation results from the fact that $\partial_\phi E_A$ is non-zero only in first order in $T$, while $\rho_\text{D}(T)\approx\frac{G}{2G_{Q}}T^{-1}$ for small $T$. Through integration one directly recovers Eq.~\eqref{eq_Is_Omega_rho_D}.

Similarly, for the double junction we may approximate $\rho_\text{dj}(T)\approx\frac{G}{\pi G_{Q}}T^{-3/2}$ for small $T$, and we thus obtain
\begin{equation}
I_S\approx \left.I_S\right|_{\Omega=2\Delta}+\frac{2eG}{\pi G_{Q}}\frac{\partial^2 E_A}{\partial\phi\partial T}(0) \left.n_A(0)^\text{st}\right|_{\Omega=2\Delta}\int_0^{\delta T}dT \sqrt{T}\ .
\end{equation}
Carrying out the integration, we arrive at Eq.~\eqref{eq_Is_Omega_rho_dj}.

\bibliographystyle{apsrev}
\bibliography{bib_superconducting}

\end{document}